\documentclass[final,5p,times,twocolumn]{elsarticle}

\usepackage[utf8]{inputenc}
\usepackage[T1]{fontenc}
\usepackage{graphicx}
\usepackage{array,subfigure}
\usepackage{amsmath,amssymb}
\usepackage{xcolor,soul,framed}
\usepackage{xfrac}
\usepackage{url}
\usepackage{epstopdf}
\DeclareGraphicsExtensions{.pdf,.png,.jpg,.jpeg,.eps}
\biboptions{sort&compress}

\journal{Optics \& Laser Technology}

\makeatletter
\def\ps@pprintTitle{%
  \let\@oddhead\@empty
  \let\@evenhead\@empty
  \let\@oddfoot\@empty
  \let\@evenfoot\@empty
}
\makeatother

\begin{document}

\begin{frontmatter}

\title{Linear computation of XPM and BER in Long-Haul Optical Systems}

\author[monash]{Ravneel Prasad}
\author[monash]{Emanuele Viterbo}

\affiliation[monash]{organization={Department of Electrical and Computer Systems Engineering, Monash University},
            city={Melbourne},
            state={VIC},
            country={Australia}}
\begin{abstract}
Cross-Phase Modulation (XPM), a critical nonlinear effect in long-haul optical communication systems utilizing Wavelength Division Multiplexing (WDM), is significantly influenced by intensity fluctuations (IFs) originating from the transmitted signal and altered by chromatic dispersion. A linear model is employed to characterize the growth of intensity fluctuations along the transmission path, demonstrating that these fluctuations are sufficient to predict the spectral characteristics of XPM on an adjacent channel. A direct correlation between frequency-domain IF growth and XPM-induced phase distortions is established and analyzed. Furthermore, the impact of XPM on the bit error ratio (BER) is shown to be analytically predictable. These analytical predictions align closely with results obtained from full nonlinear simulations. Results reveal that the evolution of IFs, especially at lower frequencies, has a pronounced effect on the XPM phase fluctuation spectra and overall phase variance. Validation through simulation confirms the model's accuracy in predicting XPM-induced phase fluctuation spectra and variance under various system configurations. These findings highlight the necessity of accounting for frequency-domain IF evolution during signal propagation in order to accurately model XPM-induced impairments, offering valuable guidance for the optimization and design of advanced optical communication systems.
\end{abstract}
\begin{keyword}
Optical communications \sep
Cross-Phase Modulation \sep
intensity fluctuation \sep
chromatic dispersion \sep
phase fluctuation spectra \sep
phase variance \sep
frequency domain modeling \sep
bit error ratio \sep
wavelength division multiplexing \sep
fiber nonlinearity.
\end{keyword}
\end{frontmatter}

\section{Introduction}\label{sec:Introduction}

Optical communication systems have evolved significantly over the years to support high-bandwidth transmission, leading to the adoption of Wavelength Division Multiplexing (WDM) and Space Division Multiplexing (SDM) systems \cite{papapavlou2022toward,di2024opportunities}. However, the transmission of multiple wavelength channels through the same optical medium introduces cross-phase modulation (XPM), which significantly affects signal integrity \cite{secondini2014,zheng2019xpm}. Variations in intensity within one wavelength channel modified by chromatic dispersion (CD) cause phase fluctuations in co-propagating wavelength channels through the Kerr effect \cite{lowery2022xpm}. These phase fluctuations further degrade the signal as they are converted into intensity fluctuations (IFs) due to CD during propagation \cite{cartaxo1999cross}. Consequently, understanding and managing these nonlinear effects in multi-wavelength channel systems is crucial.

Extensive research has been conducted on cross-phase modulation (XPM) over the years. In \cite{chiang1994cross1}, the impact of different modulation frequencies was examined both theoretically and experimentally to analyze the XPM-induced phase modulation effect in two single-frequency signals: a pump signal (which induces XPM) and a probe signal. This study introduced the well-known XPM efficiency equation \cite{chiang1994cross1,chiang1996cross}, establishing the relationship between attenuation, dispersion, fiber length, and wavelength separation. The equation was later extended in \cite{chiang1996cross} to account for multiple fiber spans connected by optical amplifiers without inline distributed dispersion compensation. This extension introduced a link factor, described by a periodic sinc function, to quantify XPM efficiency in multi-span systems. Additionally, small-signal analysis in \cite{wang1992small} demonstrated that intensity fluctuations can arise from the interplay of both amplitude and phase dynamics: chromatic dispersion produces a frequency-selective response that shapes the propagation of existing intensity modulation (IM–IM transfer) while also converting phase modulation/noise into intensity fluctuations (PM–IM conversion) through dispersion-induced differential group delay. As a result, the model from \cite{chiang1994cross1, chiang1996cross} was modified to study XPM-induced intensity modulation in \cite{cartaxo1999cross,hui1999cross}. However, these models did not account for IFs arising from pulse overlapping due to CD, an issue later addressed in \cite{ho2006cross}, which considered Gaussian-shaped pulses under the assumption of perfect dispersion compensation at the end of each fiber span. Similarly, perturbation-based approaches were developed in \cite{kumar2005second, shahi2014analytical} to study XPM, incorporating first- and second-order perturbation techniques to model pulse broadening due to CD. Analytical modeling based on the assumption that nonlinear interferences in an uncompensated transmission could be treated as additive Gaussian noise led to what is widely known as the Gaussian Model \cite{poggiolini2011analytical,poggiolini2011simple,bosco2011analytical,carena2012modeling}. In \cite{liang2014digital, liang2014analytical}, a first-order perturbation theory was applied to non-Gaussian pulses, approximating them through a summation of moving Gaussian pulses. Moreover, \cite{dar2013properties} demonstrated that phase variance due to XPM is modulation format-dependent, highlighting a limitation of the Gaussian model. Addressing this limitation, an Enhanced Gaussian Model was proposed \cite{carena2014egn}. Furthermore, the fundamental nonlinear dynamics and wave interactions associated with XPM in optical systems continue to be an active area of theoretical investigation, with recent studies providing advanced physical perspectives on nonlinear propagation phenomena \cite{zeng2024spontaneous,zeng2023solitons,zeng2021stable}.

Beyond these theoretical models, experimental studies have shown that XPM-induced distortions depend on the symbol rate \cite{hui1999cross,vassilieva2008symbol,bononi2009cross}, driven by evolving IFs along the fiber \cite{du2011optimizing,lowery2022xpm}. Although several explanations have been proposed regarding the optimal symbol rate, as reviewed in detail in \cite{lowery2022xpm}, both \cite{lowery2022xpm} and \cite{du2011optimizing} primarily focus on the IF spectrum to support their arguments. In these works, the authors propose the existence of an optimal symbol rate in systems with varying granularity, where granularity refers to the division of a fixed information bandwidth into smaller segments that are transmitted over multiple subcarriers. This optimal symbol rate is defined as the point where the sum of intrinsic and CD-induced IFs are minimized, reducing intensity fluctuations near the first null of the XPM efficiency equation. This fluctuation difference changes the nonlinear phase noise for different symbol rates. Although simulations demonstrate this effect, no analytical model included frequency-domain IF growth until it was proposed in \cite{Prasad2025intenstiy}. Prior approaches mainly addressed dispersion-induced temporal pulse changes \cite{shahi2014analytical,liang2014digital,liang2014analytical,dar2013properties}. To bridge this gap, this study establishes a direct link between IF growth and XPM in optical communication systems.

In our earlier work \cite{Prasad2025intenstiy}, a semi-analytical model was proposed to predict the growth of the IF spectra in the frequency domain. The model demonstrated the key spectral features of the IF that were dependent on the modulation format, symbol rate, pulse shape, and the fiber parameters. It further highlighted that the low-frequency IF growth is more pronounced for higher symbol rates than lower symbol rates. Our work \cite{Prasad2025intenstiy} also showed that the IF growth was mainly due to pulse overlapping due to CD within a subcarrier while the beating of one subcarrier with another subcarrier was not significant.

Building upon \cite{Prasad2025intenstiy}, this paper incorporates this phenomenon into a model of the spectra of XPM phase fluctuations. Notably, it underscores the importance of considering this growth within frequency-domain models, a factor not extensively addressed in prior work. The study then investigates the prediction of system performance, quantified by the BER, using the derived phase variance. This means that a linear simulation can be used to predict BER in the presence of XPM, considerably speeding the simulation for long-haul optical communications systems. The paper's structure is as follows: Section \ref{section:1} introduces the XPM model and its phase variance; Section \ref{simulationsconfig} describes the simulation environment and parameters; Section \ref{Results} presents the findings from the model and simulations; and Section \ref{Conclusion} summarizes the key outcomes.

\begin{figure*}
    \centering
    \includegraphics[scale=0.5]{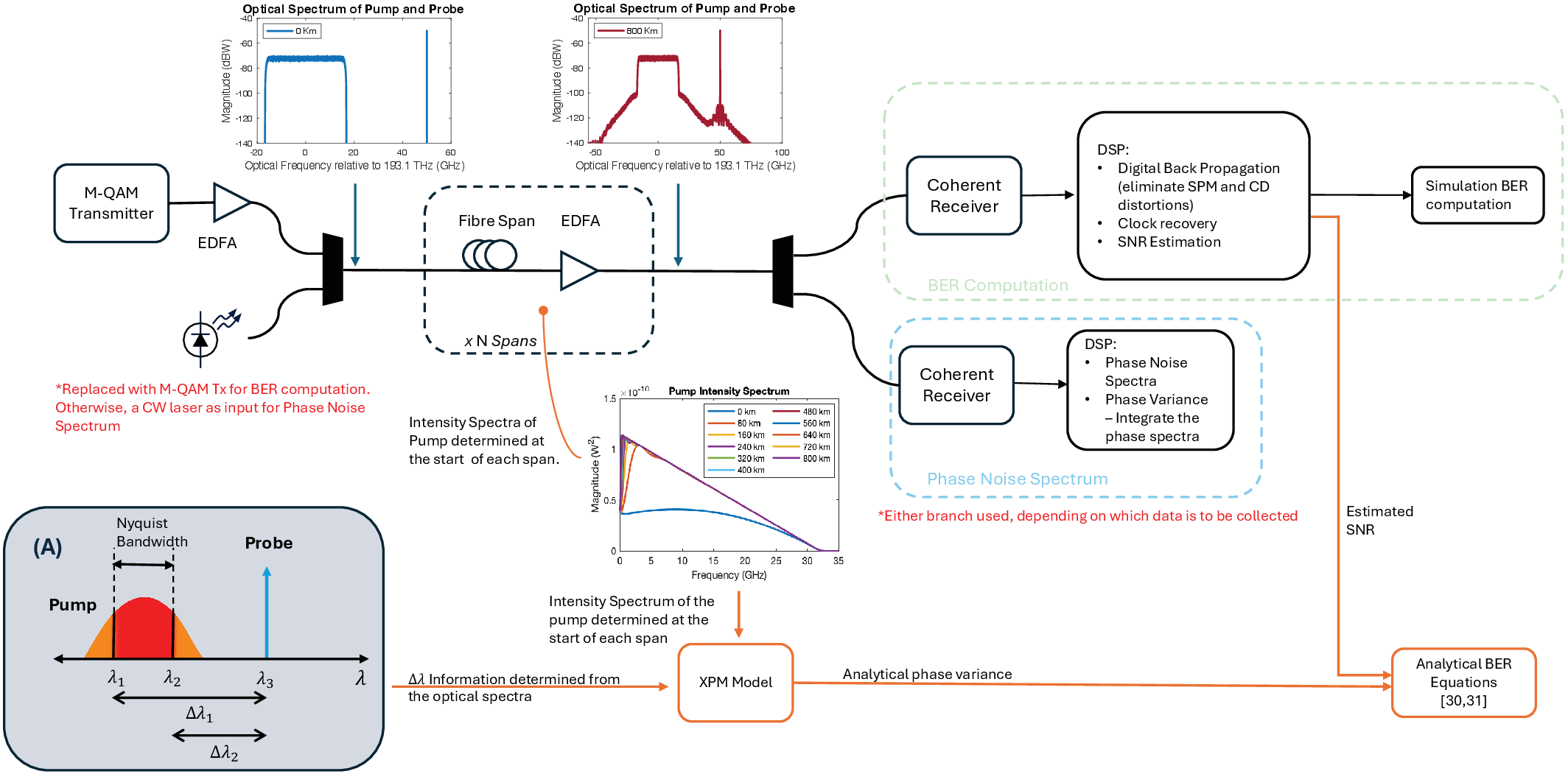}
    \caption{System overview of a pump and probe signal being transmitted along the fiber. The system also demonstrates the simulation setup used for collecting results through simulation that is indicated by black path while the orange path shows the information flow for determining analytical results. (A) A band as pump with single tone probe being transmitted along the fiber}
    \label{fig:1}
\end{figure*}

\section{XPM Model}\label{section:1}
\subsection{XPM Model - Single-tone modulation of pump}
The XPM-induced phase fluctuations caused by one signal onto a continuous wavelength (probe) can be expressed in terms of the following parameters: $D$, the dispersion parameter; $\Delta\lambda$, the wavelength separation between the pump and probe; and $L$, the length of the fiber span. The equation is given as follows for a single span fiber at any arbitrary modulation frequency as  \cite{chiang1994cross1,chiang1996cross}:

\begin{equation}\label{GeneralXPMFrequency}
    {\sigma_{{\phi_{\text{XPM}}}}(f,\Delta\lambda)} = 2\gamma {L_\text{eff}}{|P_p(f)|} {\sqrt {{\eta _{\text{XPM}} (f,\Delta\lambda)}}}\,,
\end{equation}

where, $P_p (f)$ is the pump power at a particular frequency and the effective length, $\alpha$ is the fiber attenuation, $L_\text{eff}$, is,

\begin{equation}\label{EffectiveLength}
    {L_\text{eff}} = \frac{{1 - {e^{ - \alpha L}}}}{\alpha }\,,
\end{equation}
and the XPM efficiency, $\eta_{\text{XPM}}$, at a given angular frequency $\omega = 2\pi f$ corresponding to a modulation frequency $f$, is given by:

\begin{eqnarray}\label{XPMefficiency}
    {\eta _{\text{XPM}}}(f,\Delta\lambda)  &=&\frac{{{\alpha ^2}}}{{{\omega ^2}(D \Delta\lambda)^2 + {\alpha ^2}}} \nonumber\\
    &&\times\left[ {1 + \frac{{4{{\sin }^2}(\omega {D \Delta\lambda}L/2){e^{ - \alpha L}}}}{{{{\left( {1 - {e^{ - \alpha L}}} \right)}^2}}}} \right] \,.
\end{eqnarray}

It is reasonable to assume that the intensity spectrum provides sufficient information about the intensity strength at a particular modulation frequency for characterizing the XPM effect while the optical spectrum provides the information on the wavelength separation between the pump and probe. $|P_p(f)|$ could be determined from the power spectral density (PSD) of IFs by a square root relationship since the PSD gives $|P_p(f)|^2$.

The XPM efficiency equation given in Equation (\ref{XPMefficiency}) is for a single span \cite{chiang1994cross1,chiang1996cross}. However, in long-haul system, the fiber contains $N$ spans with amplification taking place in each stage periodically compensating for the attenuation. In a dispersion unmanaged system, dispersion accumulates coherently across spans. This accumulation modifies the phasor contribution of the pump field at each span and can be modeled by the following sum:

\begin{equation}\label{vectorSum}
    \upsilon (f,\Delta\lambda) = {\sum\limits_{k = 1}^{N} {|{P_{p}^{(k)}}(f)|{e^{ - i 2\pi f {D \Delta\lambda} L (k-1)}}} } \,.
\end{equation}

The resulting XPM-induced phase shift in the probe, due to the coherent accumulation of dispersion-altered pump components, is then given by:

\begin{equation}\label{XPMAccumulatedDispersion}
    {\sigma_{{\phi_{\text{XPM}}}}}(f,\Delta\lambda) = 2\gamma {L_\text{eff}}\sqrt {{\eta _{\text{XPM} (f,\Delta\lambda)}}}~\left|\upsilon (f,\Delta\lambda)\right|\,.
\end{equation}

The signal at this stage of the proof is sinusoidally modulated,  and if $|{P_{p}^{(k)}}|$, that is the intensity at the start of the $k$th span remains constant in each span, Equation (\ref{vectorSum}) would be simplified as:

\begin{align}
|\upsilon(f,\Delta\lambda)|
  &= \left|
       P_{p}(f)
       ~e^{-i\pi f D\Delta\lambda L (N-1)}
       \frac{\sin\!\bigl(\pi N f D\Delta\lambda\,L\bigr)}
            {\sin\!\bigl(\pi f D\Delta\lambda\,L\bigr)}
     \right|\,.
\end{align}

From the equation above, the link factor similar to \cite{chiang1996cross}, can be identified to be:

\begin{equation}
    {\eta _\text{link}}(f,\Delta\lambda) = \left|\frac{{\sin (\pi Nf {D \Delta\lambda} L )}}{{\sin (\pi f {D \Delta\lambda} L )}}\right|\,.
\end{equation}
\noindent
This link factor has a periodic sinc-like shape, which is dependent on the number of spans $N$ and the wavelength separation $\Delta\lambda$. This factor contains the peaks and nulls of that determines the frequencies at which the XPM is maximized and minimized, respectively.

\subsection{XPM Model - Pass Band Signals}
For a pass-band optical signal acting as a pump, such as a QAM signal, determining a closed form equation using the standard XPM definitions would be tedious and difficult. In such case, an approximation of XPM can be made by extending the XPM formula given in Equation (\ref{XPMAccumulatedDispersion}) by now considering not only a single wavelength difference but a range of the different wavelengths causing XPM, therefore the cross-phase modulation will be the total contribution from different wavelength components within the Nyquist bandwidth (NB) of the pump signal, It will be shown later that this integration agrees well with nonlinear simulation. Here, the NB is defined by the symbol-rate-limited Nyquist band of the pump signal. It is not enlarged to include pulse-shaping excess bandwidth, and it does not include empty guard bands between subcarriers. Consider the pump and probe set up given in (A) of Figure \ref{fig:1}, the XPM induced on the probe would be given as an integral over a range of wavelength separations, indicating the XPM-induced phase shift is the average change in phase shift over the range of wavelength separations within the Nyquist bandwidth of the pump signal. The XPM-induced phase shift on the probe signal can be expressed as:

\begin{eqnarray} \label{XPMPassBand}
    \left|{\sigma_{{\phi_{\text{XPM}}}}}^{\prime}(f,\Delta\lambda)\right| = \frac{1}{{\Delta {\lambda _1} - \Delta {\lambda _2}}} \int\limits_{\Delta {\lambda _2}}^{\Delta {\lambda _1}} {{\sigma_{{\phi_{\text{XPM}}}}}(f,\Delta\lambda)}\, {\mkern 1mu} d\Delta \lambda \,.
\end{eqnarray}

This can be extended for a pump that has multiple bands, such as a multi-subcarrier signal, by applying the same averaging principle only over the NB of each occupied subcarrier. For non-contiguous multi-subcarrier spectra, Equation~(\ref{XPMPassBand}) is therefore interpreted as an average over the occupied subcarrier NBs only. Empty guard bands between subcarriers are not included in the averaging interval, and pulse-shaping excess bandwidth is not included in the integration range used in this work. The subcarrier spacing affects the result through the locations of the occupied subcarrier NBs relative to the probe, and hence through the wavelength-dependent XPM efficiency and link factor, rather than through an artificial integration over unoccupied spectral gaps.

\subsection{Average Phase Variance}
The key performance metric for XPM impairment is the average phase variance, $\sigma_{\text{XPM}}^2$, obtained by integrating the ensemble-averaged PSD of the XPM-induced phase fluctuations over frequency:
\begin{equation}\label{eq:AvgVar}
    \sigma_{\text{XPM}}^2
    = \int \mathbb{E}\!\left[
      \big|\sigma'_{\phi_{\mathrm{XPM}}}(f,\Delta\lambda)\big|^2
      \right] df.
\end{equation}
The integrand $\mathbb{E}\!\left[\big|\sigma'_{\phi_{\mathrm{XPM}}}(f,\Delta\lambda)\big|^2\right]$ is the ensemble-averaged PSD. Deriving this term requires bridging the gap between the intractable true expectation and a deterministic, computable model. This is achieved by combining three distinct statistical relationships.

\paragraph{Final Phase Distribution ($K_{\sigma}$)}
For brevity in the following derivations, the explicit dependence on $(f,\Delta\lambda)$ is omitted for variables such as $\sigma'_{\phi_{\mathrm{XPM}}}$, $\eta_{\mathrm{XPM}}$, $\upsilon$, $a_k$, $\mu$, and $R$, unless required for clarity.

It is observed through simulations that the final integrated phase fluctuation, $\sigma'_{\phi_{\mathrm{XPM}}}$, is well-described by a Rayleigh distribution. This provides a direct relationship between the PSD (the mean-square) and the squared mean-magnitude:
\begin{equation}\label{eq:RayleighRelation}
    \mathbb{E}\!\left[
    \big|\sigma'_{\phi_{\mathrm{XPM}}}\big|^2
    \right]
    = K_{\sigma} \cdot
    \left(\mathbb{E}\!\left[
    \big|\sigma'_{\phi_{\mathrm{XPM}}}\big|
    \right]\right)^2,
    \quad \text{where } K_{\sigma} = \frac{4}{\pi}.
\end{equation}
This simplifies the problem to finding the expectation of the magnitude, $\mathbb{E}[|\sigma'_{\phi_{\mathrm{XPM}}}|]$. By swapping the expectation and integral:
\begin{align}\label{eq:ExpectationInside}
    \mathbb{E}\!\left[\big|\sigma'_{\phi_{\mathrm{XPM}}}\big|\right]
    &\approx \frac{1}{\Delta\lambda_{\text{NB}}}
    \int_{\Delta\lambda_2}^{\Delta\lambda_1}
    \mathbb{E}\!\left[\sigma_{\phi_{\mathrm{XPM}}}\right]
    d\Delta\lambda \nonumber \\
    &= \frac{2\gamma L_{\mathrm{eff}}}{\Delta\lambda_{\text{NB}}}
    \int_{\Delta\lambda_2}^{\Delta\lambda_1}
    \!\! \sqrt{\eta_{\mathrm{XPM}}}
    \,\mathbb{E}\!\left[|\upsilon|\right]
    d\Delta\lambda,
\end{align}
where $\Delta\lambda_{\text{NB}} = \Delta\lambda_1-\Delta\lambda_2$ and $\mathbb{E}\!\left[|\upsilon|\right]$ is the expectation of the random phasor sum magnitude.

\begin{figure*}[ht!]
    \centering
    \subfigure[]{
        \includegraphics[width=0.4\textwidth]{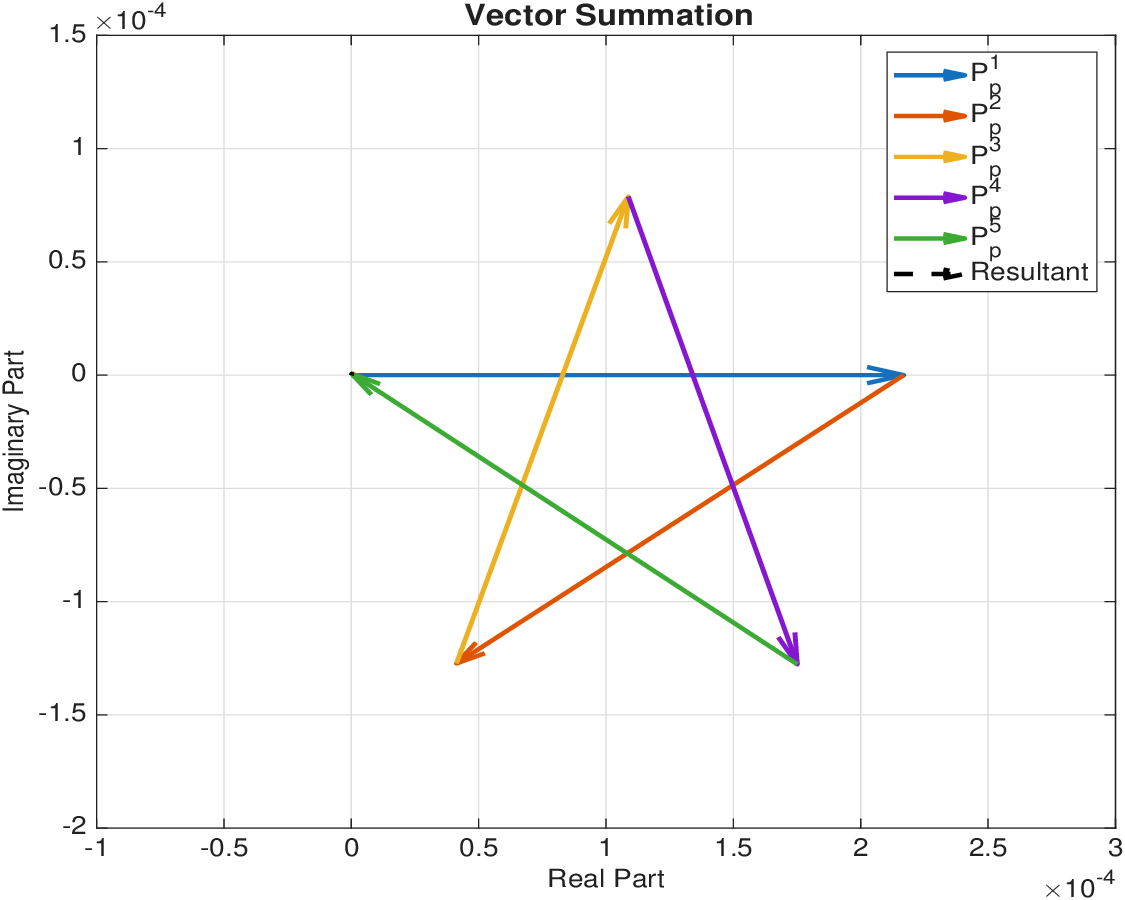}
        \label{fig:MultiSpanVectorConstant}
    }
    \subfigure[]{
        \includegraphics[width=0.4\textwidth]{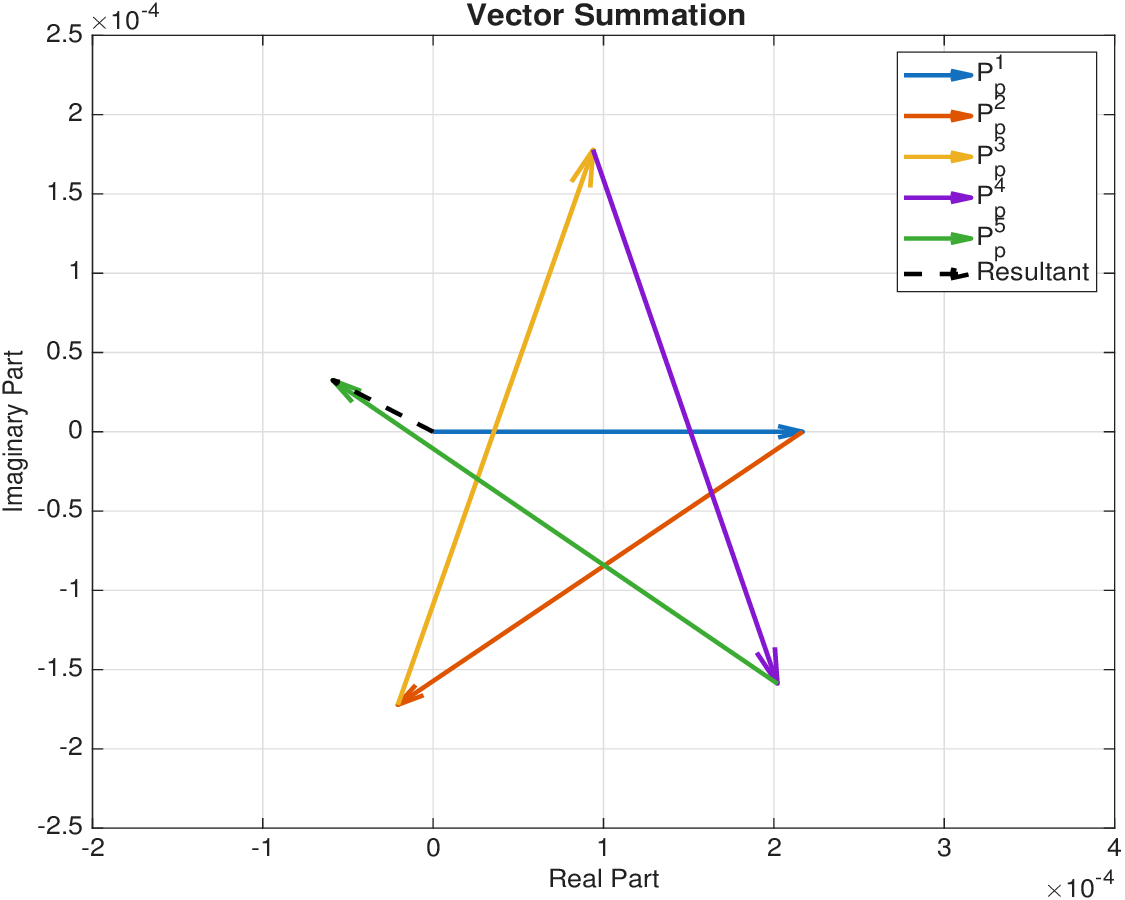}
        \label{fig:MultiSpanVectorGrowth}
    }
    \caption{Vector Summation of IFs at $\Delta\lambda=0.4$ nm and $f = 1.17$ GHz for a multi-span system. (a) The IFs of the pump signal are assumed to remain constant across spans. (b) The IFs of the pump signal evolve. The vector summation is demonstrated for the first null of the XPM efficiency curve.}
    \label{fig:MultiSpanVectorSummation}
\end{figure*}

\paragraph{Phasor Sum Ratio ($Q$)}
The term $\mathbb{E}\!\left[|\upsilon|\right]$ is related to its deterministic (mean-field) counterpart $\big|\mathbb{E}[\upsilon]\big|$ by the ratio $Q = \mathbb{E}\!\left[|\upsilon|\right] / \big|\mathbb{E}[\upsilon]\big|$.
A direct calculation of $Q$ using the true, evolving amplitude statistics (which grow from span to span, as seen in Figure~\ref{fig:MultiSpanVectorGrowth}) is mathematically intractable. To derive a \textit{tractable, closed-form expression} for this coefficient, the formal derivation in Appendix~\ref{app:bounds} relies on the simplifying assumption that the pump amplitudes $a_k$ are independent and identically distributed (i.i.d.), i.e., their statistics are constant.

It is critical to note that this i.i.d. assumption is used \textit{only} to derive the statistical factor $Q$. The final model, Eq.~(\ref{eq:ModifiedPhasor}), \textit{does} account for the span-dependent intensity growth by using the per-span $\mathbb{E}[|P_p^{(k)}(f)|^2]$. This simplified i.i.d. model allows $Q$ to be calculated for the two limiting cases of phase accumulation:

\begin{itemize}
    \item \textbf{Coherent Sum ($Q=1$):}
    When dispersion is negligible ($\phi_k \approx 0$), the phases align. The random sum becomes a scalar sum, $\mathbb{E}[|\upsilon|] = \mathbb{E}[|\sum a_k e^{-j\phi_k}|] \approx \mathbb{E}[\sum a_k] = N\mu$. The mean-field sum is $|\mathbb{E}[\upsilon]| = |\sum \mathbb{E}[a_k] e^{-j\phi_k}| \approx N\mu$. Therefore, the ratio is:
    \begin{equation}
        Q = \frac{\mathbb{E}[|\upsilon|]}{|\mathbb{E}[\upsilon]|} \approx \frac{N\mu}{N\mu} = 1.
    \end{equation}
    
    \item \textbf{Incoherent Sum ($Q \approx \sqrt{4/\pi}$):}
    When dispersion is significant, the phases $\phi_k(f,\Delta\lambda)$ decorrelate. The i.i.d. model in Appendix~\ref{app:bounds} provides a general upper bound dependent on the specific geometric factor $R(f,\Delta\lambda)^2$:
    \begin{equation}
        Q(f,\Delta\lambda) \le \sqrt{1 + \frac{N(4-\pi)}{\pi R(f,\Delta\lambda)^2}}.
    \end{equation}
    The main PSD model, however, involves the integral in Equation (\ref{eq:ExpectationInside}), which \textit{averages over the $\Delta\lambda$ range}. This integration acts as an averaging process over many phase realizations, physically justifying the approximation of the statistical factor $Q(f,\Delta\lambda)$ by its \textit{average-case} value rather than its worst-case (at $R \to 0$). Crucially, this justification holds only if the integration bandwidth ($\Delta\lambda_{\text{NB}}$) is sufficiently large. If the bandwidth is too narrow, the phase diversity is insufficient to ensure a uniform distribution, and the global average ceases to be a valid estimator. Provided the bandwidth is sufficient, the specific $R(f,\Delta\lambda)^2$ in the bound is replaced with its average value, $\langle R^2 \rangle = N$ (derived in Appendix~\ref{app:bounds}). This yields the \textit{average-case upper bound}:
    \begin{equation}
        Q \approx \sqrt{1 + \frac{N(4-\pi)}{\pi N}} = \sqrt{1 + \frac{4-\pi}{\pi}} = \sqrt{\frac{4}{\pi}}.
    \end{equation}
    This average-case bound is used as the approximation for $Q$ in the incoherent regime.
\end{itemize}

\paragraph{Amplitude Statistics ($K_a$)}
The mean-field sum $\big|\mathbb{E}[\upsilon]\big|$ is based on the mean amplitude $\mathbb{E}[a_k]$. The amplitudes $a_k = |P_p^{(k)}(f)|$ are assumed to be Rayleigh distributed, which provides a fixed ratio between the mean and the root mean square (RMS), $\sqrt{\mathbb{E}[a_k^2]}$:
\begin{equation}\label{eq:RayleighMeanPower}
    \mathbb{E}[a_k]
    = \sqrt{\frac{\pi}{4}}\,
      \sqrt{\mathbb{E}[a_k^2]}
    = K_a \cdot \sqrt{\mathbb{E}[a_k^2]}.
\end{equation}
Therefore, $\big|\mathbb{E}[\upsilon]\big| = \big|\sum \mathbb{E}[a_k] e^{-j\phi_k}\big| = \big|\sum (K_a \sqrt{\mathbb{E}[a_k^2]}) e^{-j\phi_k}\big|$.

\paragraph{Final Combined Equation}
We now substitute all these relations back into \eqref{eq:RayleighRelation}. Factors $Q$ and $K_a$ can be factored out of the integral:
\begin{align}
    \mathbb{E}\!\left[\big|\sigma'_{\phi_{\mathrm{XPM}}}\big|^2\right]
    &\approx K_{\sigma} \left(
      \frac{2\gamma L_{\mathrm{eff}}}{\Delta\lambda_{\text{NB}}}
      \int \! \sqrt{\eta}
      \,\mathbb{E}\!\left[|\upsilon|\right]
      d\Delta\lambda
    \right)^2 \nonumber \\
    &\approx K_{\sigma} \left(
      \frac{2\gamma L_{\mathrm{eff}}}{\Delta\lambda_{\text{NB}}}
      \int \! \sqrt{\eta}
      \,(Q \cdot \big|\mathbb{E}[\upsilon]\big|)
      d\Delta\lambda
    \right)^2 \nonumber \\
    &\approx K_{\sigma} \left(
      \frac{2\gamma L_{\mathrm{eff}}}{\Delta\lambda_{\text{NB}}}
      \int \!\sqrt{\eta}
      \,(Q \cdot K_a \cdot |\upsilon'|)
      d\Delta\lambda
    \right)^2 \nonumber \\
    &\approx \underbrace{K_{\sigma} (Q \cdot K_a)^2}_{K}\times\nonumber\\
      &\quad\left|
      \frac{2\gamma L_{\mathrm{eff}}}{\Delta\lambda_{\text{NB}}}
      \int_{\Delta\lambda_2}^{\Delta\lambda_1} \! \sqrt{\eta(f,\Delta\lambda)}
      \,|\upsilon'(f,\Delta\lambda)|
      d\Delta\lambda
      \right|^2,
\end{align}
where $|\upsilon'(f,\Delta\lambda)|$ is the deterministic phasor sum based on the \textit{RMS amplitudes}:
\begin{equation}\label{eq:ModifiedPhasor}
    |\upsilon'(f,\Delta\lambda)|
    = \left|
      \sum_{k=1}^{N}
      \sqrt{\mathbb{E}[|P_p^{(k)}(f)|^2]}\,
      e^{-j2\pi f D\Delta\lambda L (k-1)}
      \right|.
\end{equation} 

where $\mathbb{E}[|P_p^{(k)}(f)|^2]$ denotes the span-dependent average power spectral density (PSD) of the pump intensity fluctuations at the input of the $k$th span. This quantity can be obtained numerically by linearly propagating the pump waveform to the input of each span and computing its IF spectrum, or analytically using the IF-growth model developed in \cite{Prasad2025intenstiy}. The detailed derivation of the IF spectra for unshaped signals is provided in \cite{Prasad2025intenstiy}. The present paper focuses mainly on unshaped signals; however, the same XPM framework can be applied to shaped signals once the corresponding span-dependent IF spectra are known. In shaped systems, the shaping distribution modifies the IF spectrum, and therefore changes the resulting XPM phase-fluctuation spectrum, phase variance, and BER, while the overall procedure of using the span-dependent IF spectra as inputs to the XPM model remains unchanged. The IF spectra of shaped signals and their impact on XPM are considered separately in \cite{prasad2026intensity}.

This defines the overall statistical factor $K$:
\begin{equation}
    K = K_{\sigma} (Q \cdot K_a)^2 = \frac{4}{\pi} \left( Q \cdot \sqrt{\frac{\pi}{4}} \right)^2 = Q^2.
\end{equation}
This result elegantly summarizes the statistical impact based on the simplified i.i.d. model:
\begin{itemize}
    \item \textbf{Coherent Case ($Q=1$):} $K = 1^2 = \mathbf{1}$.
    \item \textbf{Incoherent Case ($Q \approx \sqrt{4/\pi}$):} $K = (\sqrt{4/\pi})^2 = \mathbf{4/\pi}$.
\end{itemize}
In this work, these two values are used as limiting cases rather than as a fitted transition law. The coherent value $K=1$ is appropriate when the span-to-span phasor rotation $2\pi fD\Delta\lambda L$ remains small over the frequency and wavelength range that dominates the integral, so that the XPM contributions add nearly coherently. The average incoherent value $K=4/\pi$ is appropriate when dispersion, wavelength separation, bandwidth, and span count provide sufficient phase diversity for the geometric factor to be well represented by its average value. For intermediate cases, the same statistical framework can be used to evaluate a system-dependent value of $K$: the distribution of the phasor-sum ratio $Q=\mathbb{E}[|\upsilon|]/|\mathbb{E}[\upsilon]|$ may be estimated for the relevant dispersion, bandwidth, span count, subcarrier structure, and modulation statistics, after which $K=K_{\sigma}(QK_a)^2$ can be computed. Thus, the present derivation provides the coherent and incoherent bounds and the procedure for obtaining an intermediate value, rather than imposing a universal smooth interpolation. This is important because the intermediate distribution itself changes with dispersion, signal bandwidth, symbol distribution, and spectral structure. In practice, $K=1$ should be used as a conservative choice for near-zero walk-off or very narrow pump Nyquist bandwidths, whereas $K=4/\pi$ is suitable for the SSMF long-haul cases considered here, where the integration over pump Nyquist bandwidth and the accumulated dispersion provide strong phase averaging.
The final model for the total average XPM-induced phase variance is:
\begin{equation}\label{eq:AverageXPMPassBand}
\begin{split}
    \mathbb{E}\!\left[
    \big|\sigma'_{\phi_{\mathrm{XPM}}}\big|^2
    \right]
     &\approx K\!\left|
       \frac{2\gamma L_{\mathrm{eff}}}
            {\Delta\lambda_1-\Delta\lambda_2}
       \right|^2  \\
     &\quad\times
       \left|
       \int_{\Delta\lambda_2}^{\Delta\lambda_1}
         \sqrt{\eta_{\mathrm{XPM}}(f,\Delta\lambda)}
         \,|\upsilon'(f,\Delta\lambda)|
       d\Delta\lambda
       \right|^2 \,.
    \end{split}
\end{equation}

Equation~(\ref{eq:AverageXPMPassBand}) is a tractable model where $K=1$ is used for coherent accumulation (negligible phase change in vector summation) and $K=4/\pi$ is used for the average-case incoherent accumulation. The value of $K$ can be adjusted based on the degree of coherence expected in the system, which is influenced by parameters such as dispersion, span length, wavelength separation and number of spans where $K$ is in the range of $[1, 4/\pi]$. In this work, it is found that for typical standard single mode fiber (SSMF), the incoherent case with $K\approx 4/\pi$ provides a good approximation. In this work, it is found that using $K\approx 4/\pi$ provides good agreement with nonlinear simulations for standard single mode fiber (SSMF) systems with significant dispersion over multiple spans.

\subsection{Vector Summation Representation of XPM}
\subsubsection{Multi-span - Fixed $\Delta\lambda$ and $f$}
\begin{figure}
    \centering
    \includegraphics[height=6cm, width=8cm]{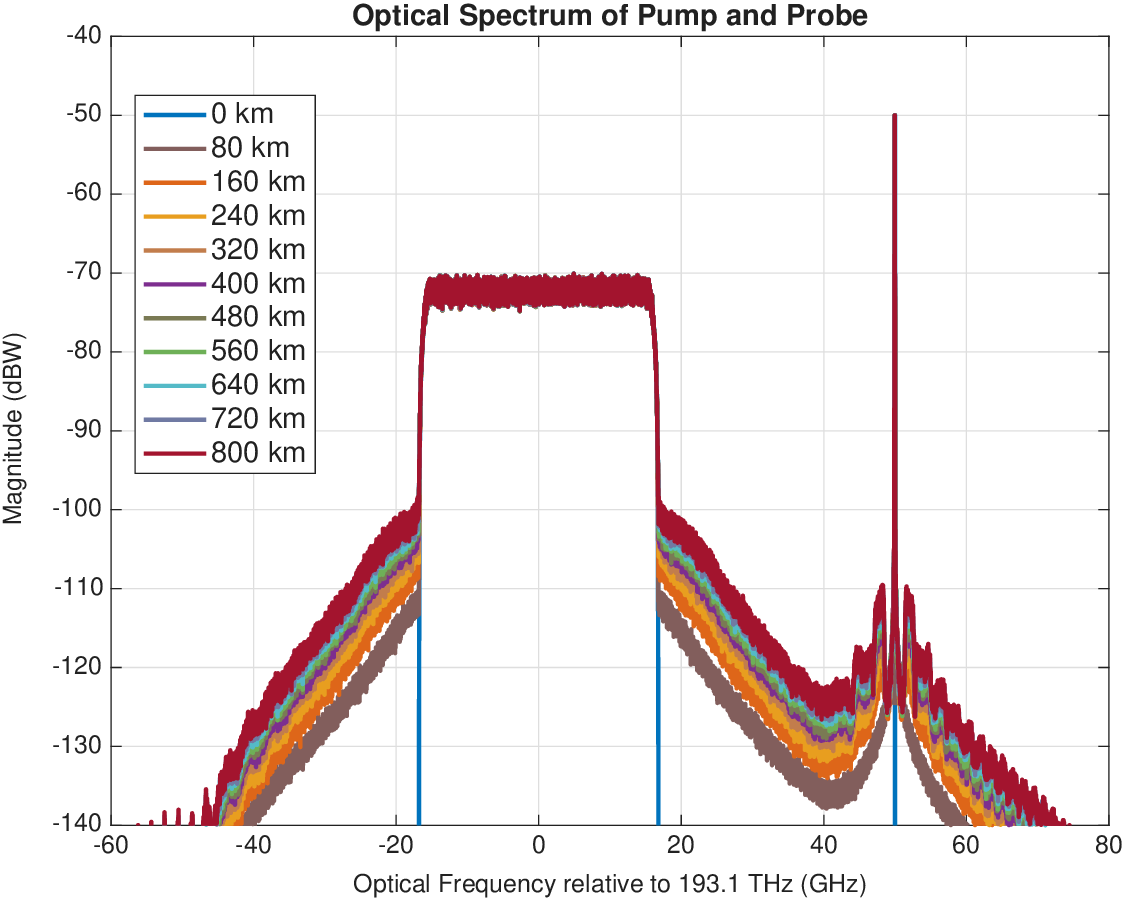}
    \caption{Single 16-QAM pump signal that is spaced $50$ GHz apart from the probe.}
    \label{fig:OpticalSpectraPumpProbeOneSub}
\end{figure}

In multi-span systems, the amplifiers at the beginning of each span compensate for the fiber's attenuation, which makes the growth of intensity fluctuations (IF) due to chromatic dispersion (CD) more pronounced over long distances. The total XPM is evaluated as a coherent vector summation of IF contributions from each span, as described by Equation (\ref{eq:ExpectationInside}) for fixed values of $f$ and $\Delta\lambda$. Figure \ref{fig:MultiSpanVectorSummation} illustrates this vector summation for a 5-span (400 km) system. The comparison is performed for a fixed wavelength separation of $\Delta \lambda = 0.4$ nm and a modulation frequency of $1.17$ GHz. The simulation uses the single-subcarrier pump and probe optical spectra shown in Figure \ref{fig:OpticalSpectraPumpProbeOneSub}. Two distinct models are demonstrated in Figure \ref{fig:MultiSpanVectorSummation}. The first case, (a), illustrates a simplified \textit{Constant IF Model}. Here, the IFs of the pump signal are assumed to remain constant, meaning the IF spectrum at the \textit{input} of every span is assumed to be identical to the spectrum at the transmitter (i.e., $|P_p^{(k)}(f)| = |P_p^{(1)}(f)|$ for all $k$). In contrast, the second case, (b), shows the more physically accurate \textit{IF Growth Model}. In this model, the IFs of the pump signal are allowed to evolve, and the IF spectrum at the input of the $k$-th span, $|P_p^{(k)}(f)|$, differs for each span as it grows along the fiber. This comparison highlights that XPM evaluation should not rely on the simplified constant IF assumption. The resulting XPM contribution becomes significantly stronger when the CD-induced IF growth is properly accounted for.

\begin{figure}[]
    \centering
    \includegraphics[height=6cm, width=8cm]{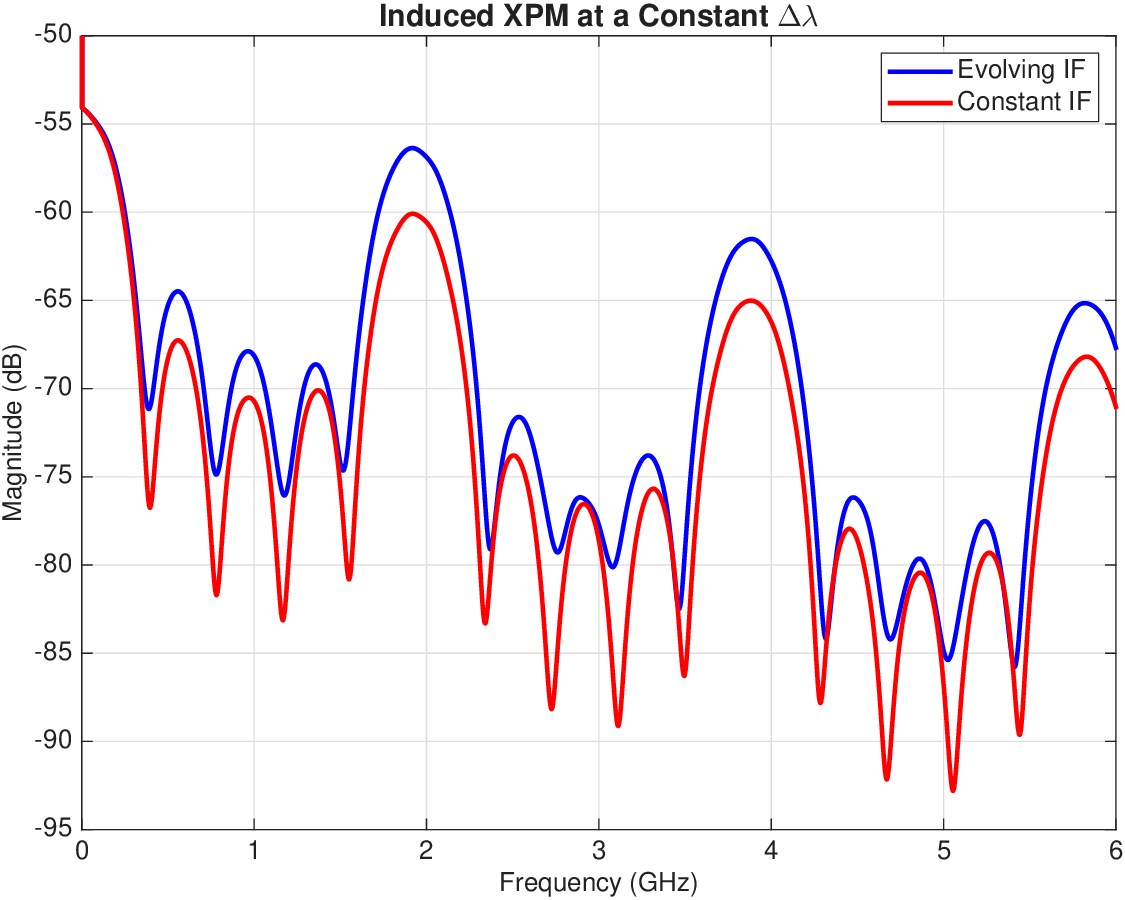}
    \caption{The shape of the link factor for a multi-span system with and without consideration of IF evolution in Equation~(\ref{eq:AverageXPMPassBand}) at a fixed wavelength separation.}
    \label{fig:MultiSpanXPM}
\end{figure}

\subsubsection{Modified XPM Link Factor - Fixed $\Delta\lambda$}
The XPM phase fluctuation model in Equation (\ref{eq:AverageXPMPassBand}) depends on the IF spectral amplitude, $|P_p^{(k)}(f)|$, at the start of each span $k$. A simplified approach, illustrated in Figure \ref{fig:MultiSpanVectorConstant}, assumes a \textit{constant IF} model. This model assumes the IF spectrum at the \textit{input} of every span is identical to the transmitter's (i.e., $|P_p^{(k)}(f)| = |P_p^{(1)}(f)|$ for all $k$), neglecting any evolution of the IF spectrum during propagation. However, the pump's IF spectrum is known to evolve and grow as it propagates through the fiber. This section analyzes a more accurate model that accounts for this IF growth, as illustrated in Figure \ref{fig:MultiSpanVectorGrowth}. In this model, the vector summation of Equation (\ref{eq:ModifiedPhasor}) uses the evolving IF amplitude spectrum, $|P_p^{(k)}(f)|$, which is the square root of the IF PSD. By accounting for this IF growth, the magnitude of the terms in the vector sum increases with $k$, resulting in a significantly stronger total XPM component compared to the constant IF model. The resultant XPM component for different modulation frequencies at a fixed $\Delta\lambda$ of $0.4\,$nm is shown in Figure \ref{fig:MultiSpanXPM}. This growth in the IF spectrum's amplitude across spans alters the coherent vector summation, increasing the XPM components at most frequencies. This effect causes the nulls of the multispan XPM response (typically described by the periodic sinc link factor) to become shallower, and the subsequent peaks to become stronger, except for the peak near the DC frequencies.
\section{Simulation}\label{simulationsconfig}
The analytical model was validated against numerical simulations performed using VPItransmissionMaker Optical Systems v11.5 to validate the accuracy of the model. The system model simulated is shown in Figure \ref{fig:1}, where a continuous wave laser was used as a probe to capture the cross-phase modulation induced by the pump. The IF spectra modelled using the expressions given in \cite{Prasad2025intenstiy} of the pump at the start of each span is used in the XPM model described in Section \ref{section:1}. A band-pass pump signal was generated using $2^{18}$ random symbols that were modulated using a 16-QAM modulation scheme. The simulation was averaged over 50 independent runs to obtain a smooth average spectrum, corresponding to a total of $50 \times 2^{18}$ symbols.

\begin{figure}[t]
    \centering
    \includegraphics[height=6cm, width=8cm]{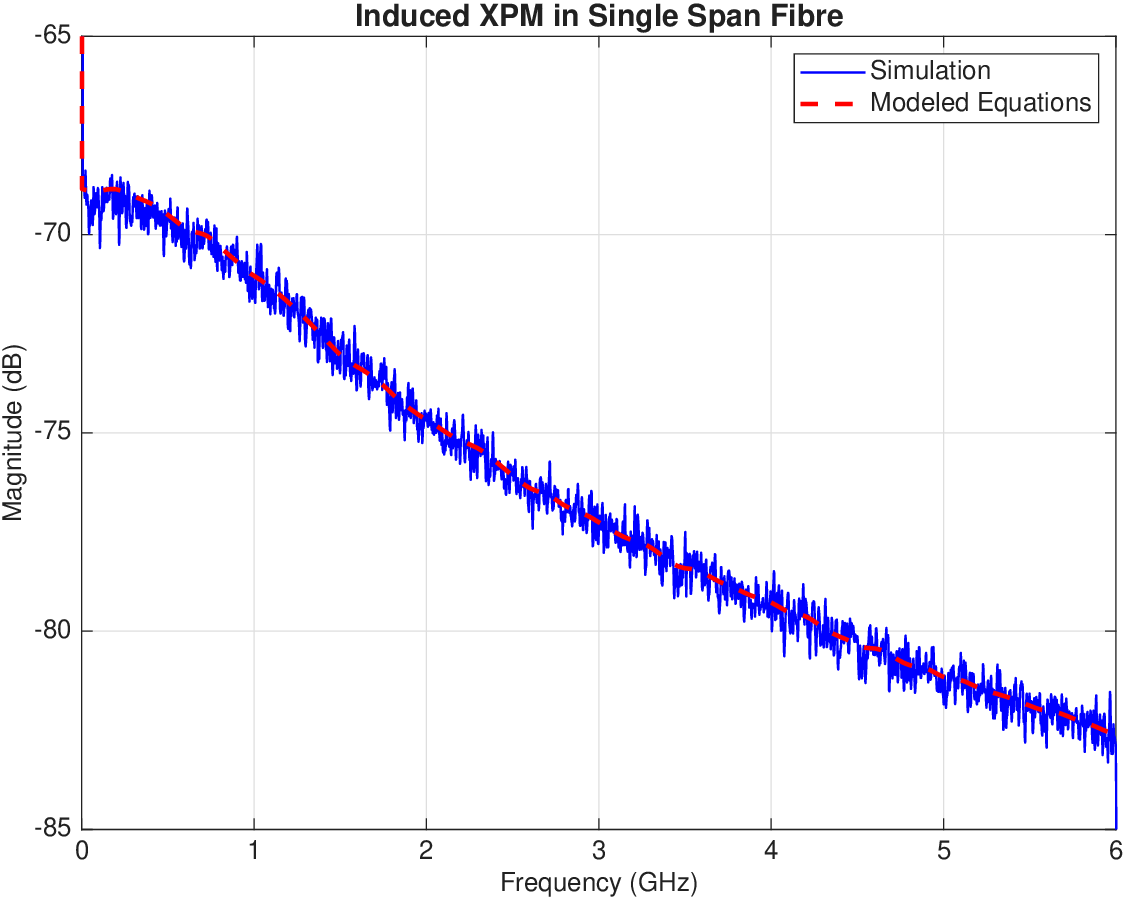}
    \caption{XPM induced phase fluctuation spectrum on the probe from a single subcarrier pump in a single-span system. The analytical model is evaluated using Equation~(\ref{eq:AverageXPMPassBand}).}
    \label{fig:SingleSpanXPMModelVPI}
\end{figure}

The simulated system was evaluated using a \textit{split-step algorithm} for the nonlinear system using the following parameters: attenuation, $\alpha$, of $0.04605$ Np/km (0.2 dB/km), dispersion, $D$, of $16$ ps/nm/km, nonlinear index, $n_2$, of $2.6 \times 10^{-20}$ m$^2$/W and effective fiber cross section area, $A_{eff}$, $80$ $\mu$m$^2$. Each fiber span consisted of an 80 km long fiber. The pump signal was set to $1552.52$ nm with a bandwidth of $32$ GHz and a launch power of $1$ mW unless stated otherwise, while the probe signal was set to $10$ $\mu$W and $50$ GHz apart from the pump.

To ensure the phase distortions retrieved from the simulation were purely from the XPM induced on the probe was filtered using a 12 GHz bandpass filter to isolate the probe signal and exclude the pump's sidebands.

To evaluate the XPM-induced phase fluctuation spectra (Equation \ref{eq:AverageXPMPassBand}) and phase variance (Equation \ref{eq:AvgVar}) described in Section \ref{section:1}, MATLAB simulations were performed for both single-span and multi-span systems. The XPM component induced on the probe by the pump signal was calculated and validated against the VPI simulation results.

\section{Results}\label{Results}
\subsection{Single-Span XPM Model}
The analytical model is first validated for a single-span system. For this case, the XPM-induced phase fluctuation spectrum is calculated using Equation~(\ref{eq:AverageXPMPassBand}) with $N=1$ and $K=1$. This calculation determines the cross-phase induced by the pump by integrating the contributions over a range of $\Delta\lambda$ corresponding to the pump signal's NB. Figure~\ref{fig:SingleSpanXPMModelVPI} compares the output of this analytical model against a full VPI simulation for a single 80 km span. The results demonstrate good agreement, validating the model's application to a single-span. The spectrum exhibits the well-known low-pass characteristic, as demonstrated in \cite{chiang1994cross1,chiang1996cross} for a sinusoidally modulated pump at a fixed wavelength difference. However, for the pass-band pump signal, this low-pass characteristic is slightly varied, as the model correctly integrates the XPM efficiency over a range of $\Delta\lambda$.

\begin{figure}[t]
    \centering
    \includegraphics[height=6cm, width=8cm]{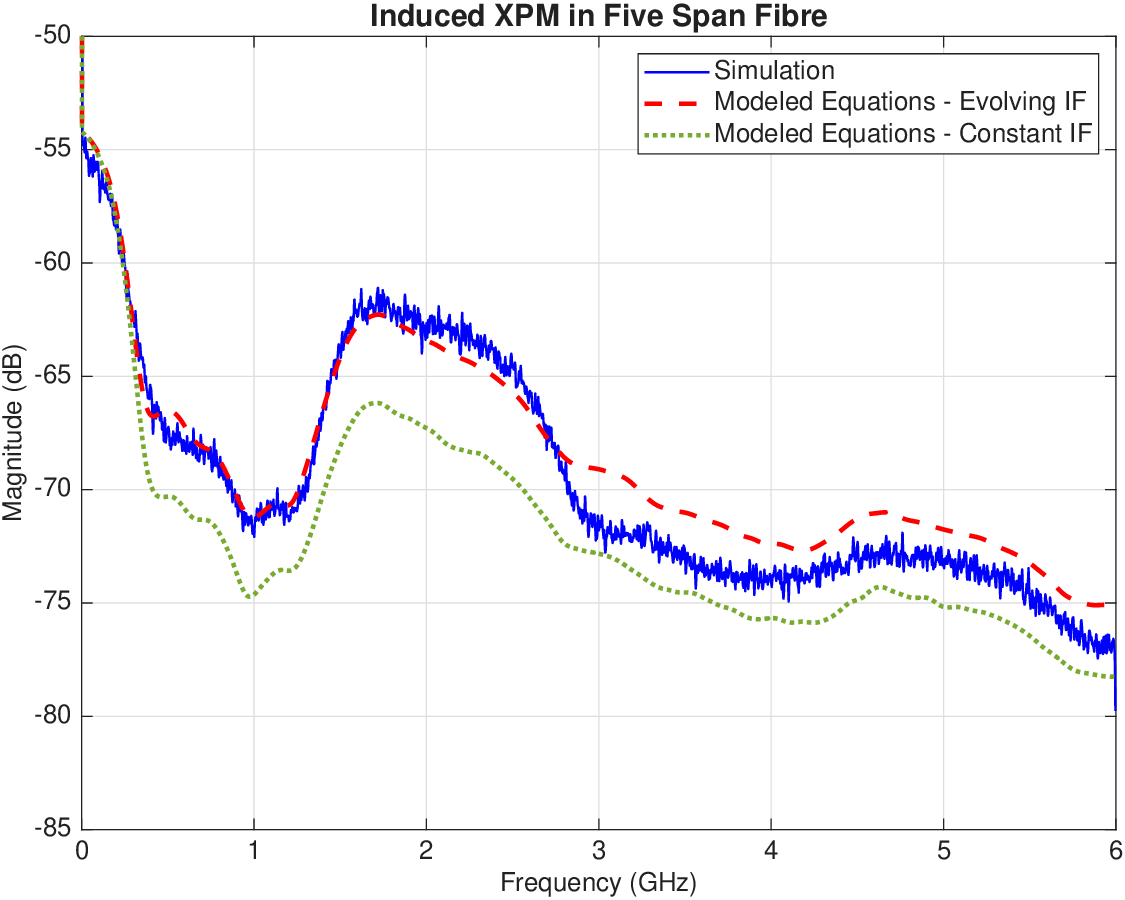}
    \caption{XPMinduced phase fluctuation spectrum on the probe for a single subcarrier pump in a 5 span system. Neglecting IF growth leads to significant underestimation of the XPM component. The analytical model incorporates IF evolution using Equation~(\ref{eq:AverageXPMPassBand}).}
    \label{fig:FiveSpanXPMModelVPI}
\end{figure}

\subsection{Multi-Span - XPM Model}
For multi-span systems ($N>1$), the total XPM-induced phase noise spectrum is evaluated by the integrand of Equation~(\ref{eq:AvgVar}), which coherently sums the IF contributions from each span and integrates over the pump's pass-band.

\subsubsection{Effect of IF Growth on the XPM Link Factor}
The spectrum of the total phase noise in a multispan system is strongly influenced by the evolution of the IFs due to CD. Figure~\ref{fig:MultiSpanXPM} illustrates the XPM spectrum at a fixed $\Delta\lambda$, comparing a "Constant IF" model (where IFs are assumed to be the same at each span) with an "Evolving IF" model (which accounts for IF growth). The growth in IFs results in the nulls of the multispan XPM (described by the link factor) to be shallower and the peaks stronger, except for the one near DC.

\subsubsection{Validation of the Full Spectral Model}
This effect is confirmed in Figure~\ref{fig:FiveSpanXPMModelVPI}, which shows the full XPM spectrum for a 5-span system. The VPI simulation (blue, solid line) is compared against two analytical models:
\begin{itemize}
    \item The "Constant IF" model (green, dotted line), which neglects IF growth, results in a significant underestimation of the XPM component.
    \item The "Evolving IF" model (red, dashed line), which incorporates IF growth, provides a much better approximation and aligns closely with the VPI simulation.
\end{itemize}
This highlights that consideration of evolving IFs is critical for the accurate evaluation of XPM. The integration over a range of $\Delta\lambda$ in Equation~(\ref{eq:AvgVar}) also explains why the sharp nulls seen in Figure~\ref{fig:MultiSpanXPM} (at a fixed $\Delta\lambda$) are "smeared out" into broader peaks in the final spectrum.

\begin{figure}
    \centering
    \includegraphics[height=6cm, width=8cm]{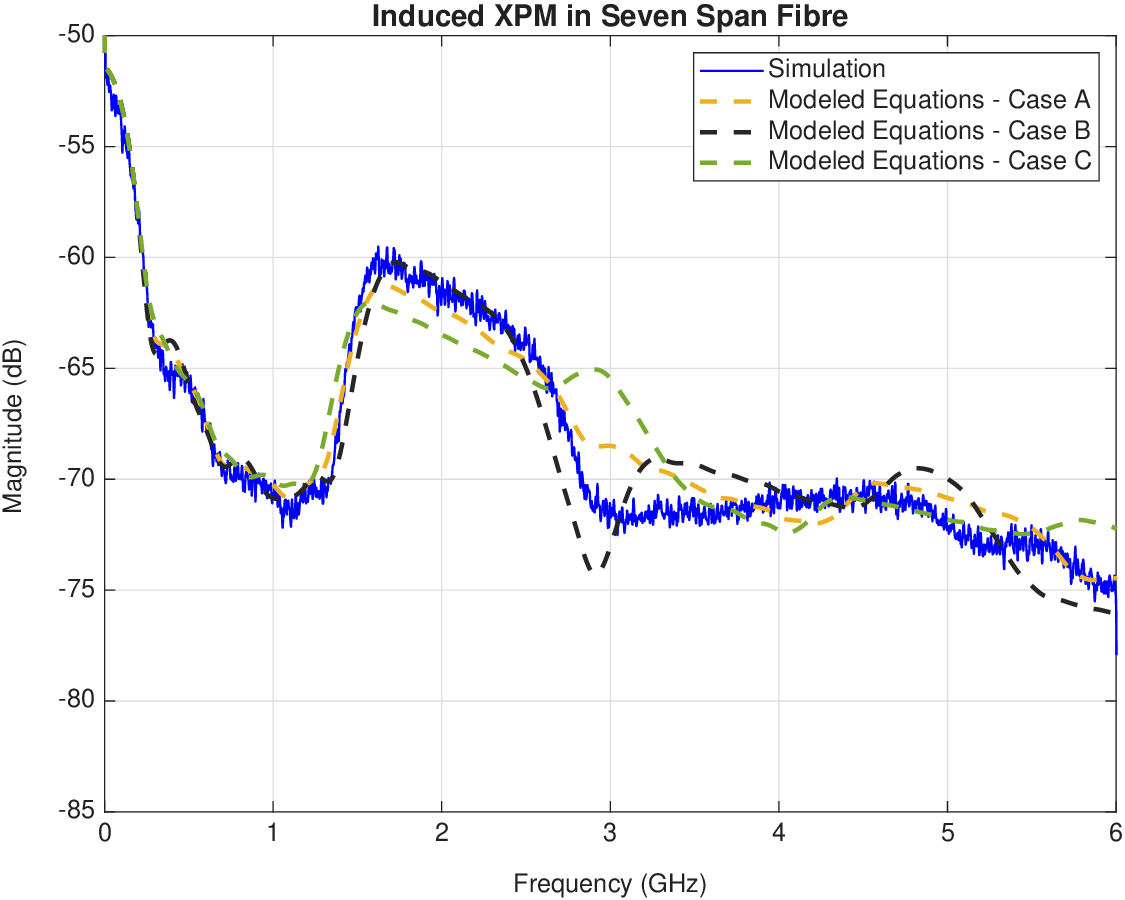}
    \caption{The importance of considering the Nyquist bandwidth of the pump signal in evaluating the $\Delta\lambda$ for 7 span system where the pump is a single subcarrier signal. There are three cases evaluated using modeled Equation~(\ref{eq:AverageXPMPassBand}): Case A: integration over the full Nyquist bandwidth of the pump (i.e., $\Delta\lambda_1 = 0.5304$ nm, $\Delta\lambda_2 = 0.2733$ nm); Case B: integration over a narrower bandwidth within the pump NB ($\Delta\lambda_1 = 0.5056$ nm, $\Delta\lambda_2 = 0.2988$ nm); Case C: integration extended beyond the pump NB ($\Delta\lambda_1 = 0.5554$ nm, $\Delta\lambda_2 = 0.2483$ nm).}
    \label{fig:BWConsideration}
\end{figure}

\begin{figure}[h!]
    \centering
    \includegraphics[height=6cm, width=8cm]{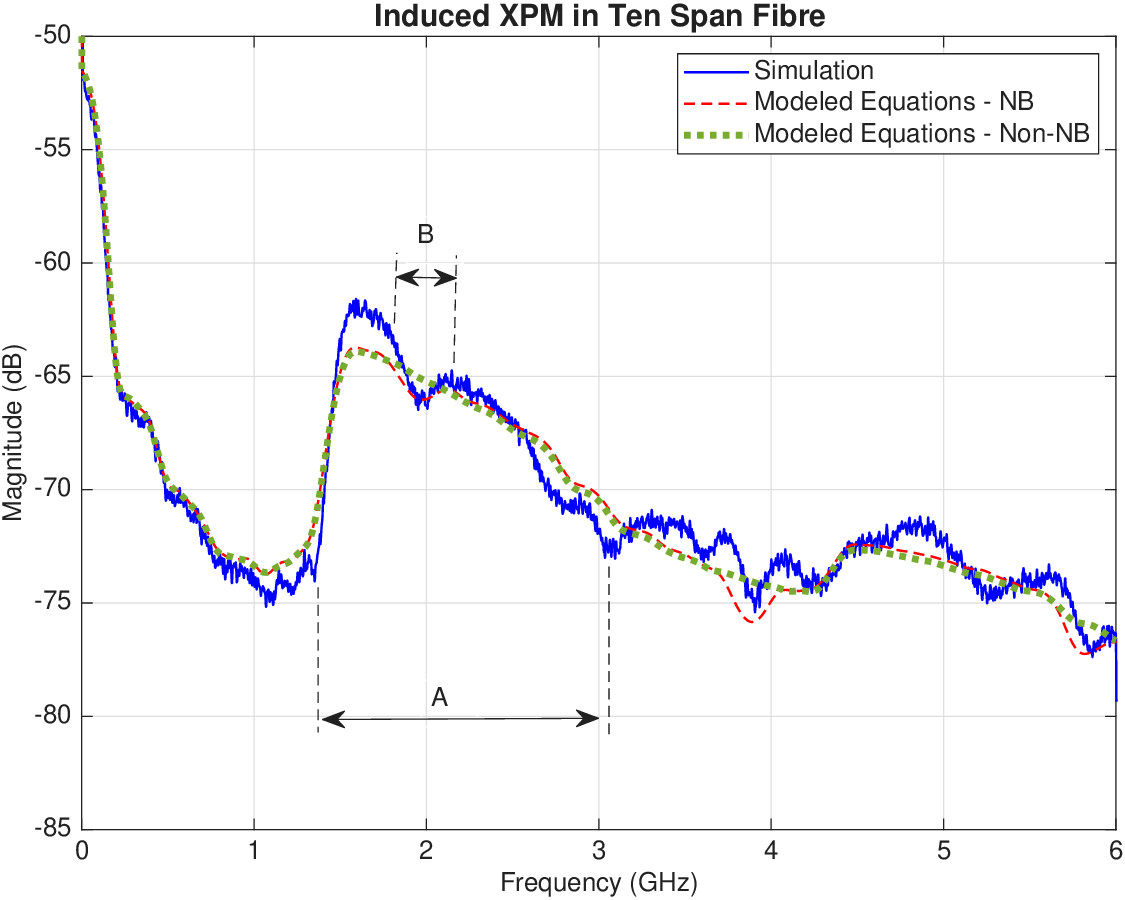}
    \caption{Two subcarrier pump signal induced Phase noise spectrum on to a probe in a 10 span system evaluated using Equation~(\ref{eq:AverageXPMPassBand}) and simulation, where (a) indicates the XPM component due to the second major peak contributions of the sinc like function representing the link factor while (b) indicates the dip in the XPM spectra due to spacing between the subcarriers.}
    \label{fig:TenSpanXPM}
\end{figure}

\subsubsection{Justification of Integration Bounds}
A key aspect of the model is the selection of the integration range \([\Delta\lambda_2, \Delta\lambda_1]\). This range is not arbitrary or a "curve fitting" parameter; it is physically determined by the pump's Nyquist bandwidth (NB), which contains the intensity fluctuations responsible for XPM. Figure~\ref{fig:BWConsideration} illustrates this for a 7-span system by comparing three integration ranges:
\begin{itemize}
    \item \textbf{Case A:} Integration over the full $32~\mathrm{GHz}$ Nyquist bandwidth of the pump. The resulting spectrum closely aligns with the simulated behavior.
    \item \textbf{Case B:} Integration over a narrower $26~\mathrm{GHz}$ bandwidth. This underestimates the width of the XPM spectral peaks.
    \item \textbf{Case C:} Integration extended beyond the pump NB to $38~\mathrm{GHz}$. This produces an overly widened spectrum that does not match the simulation.
\end{itemize}
This comparison validates that confining the integration to the pump's NB (Case A) is the physically correct approach. This point is important for the interpretation of non-rectangular spectra. In this work, the integration range is deliberately restricted to the Nyquist bandwidth of the pump signal, or to the Nyquist bandwidth of each subcarrier in a multi-subcarrier pump. The excess bandwidth introduced by pulse shaping is not included. As shown by the comparison with wider integration bounds, extending the integration beyond the NB unnecessarily smooths the spectral features and degrades the agreement with simulation.

\begin{figure*}[ht]
    \centering
    \subfigure[]{
        \includegraphics[width=0.4\textwidth]{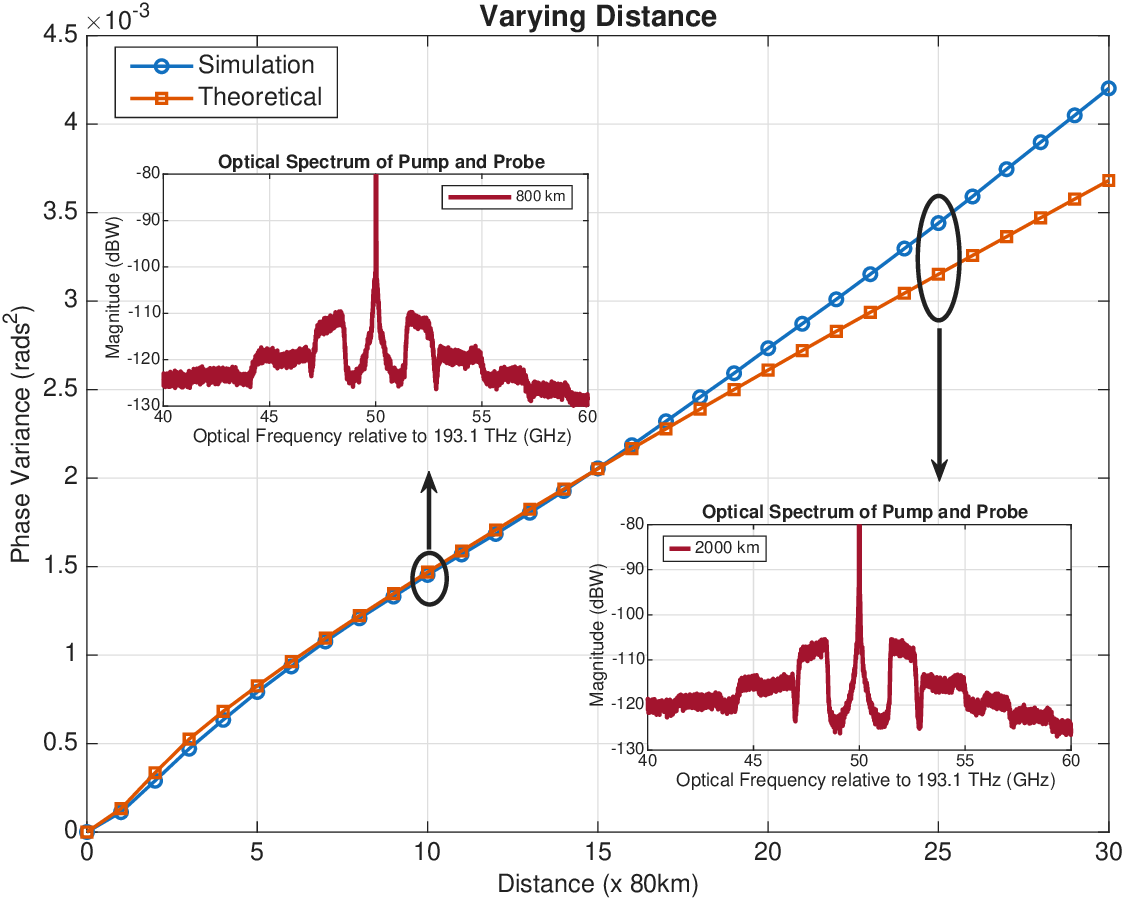}
        \label{fig:VaryingDistance}
    }
    \subfigure[]{
        \includegraphics[width=0.4\textwidth]{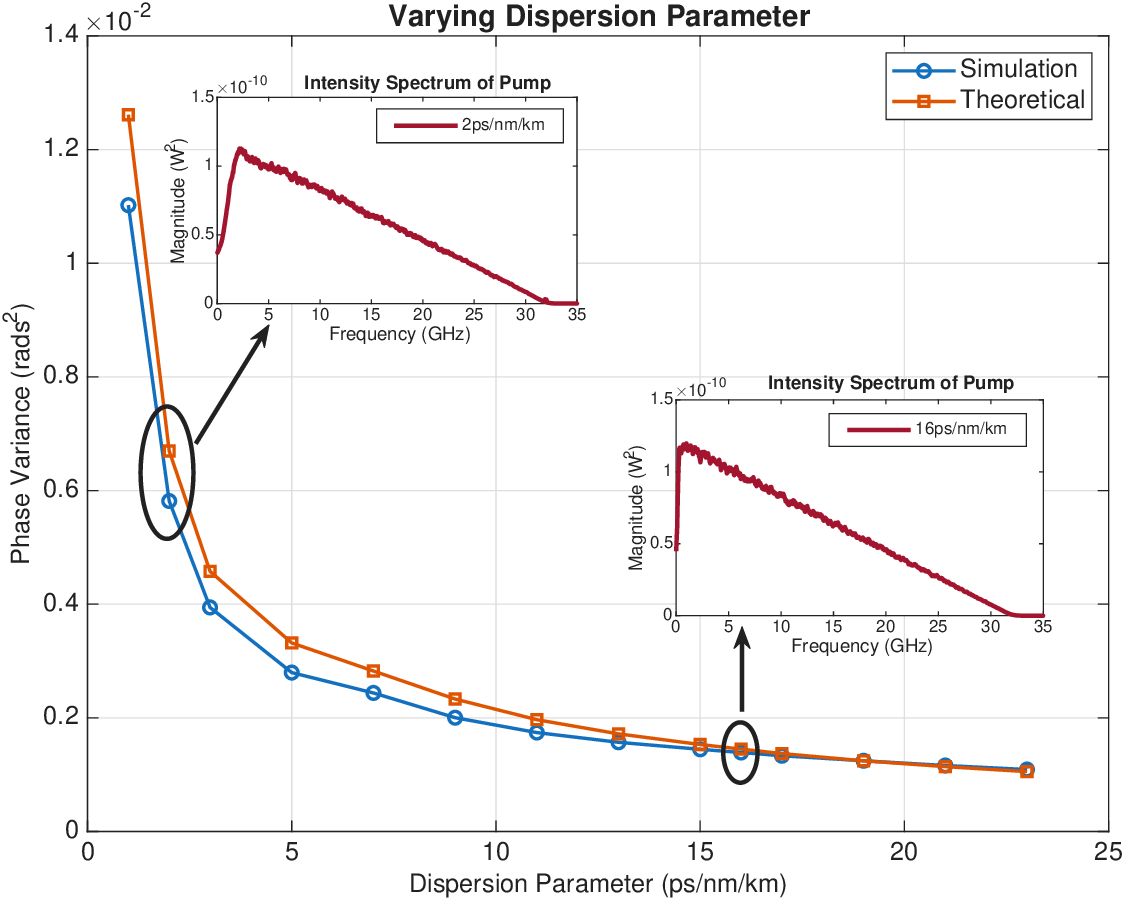}
        \label{fig:VaryingDispersion}
    }
    \subfigure[]{
        \includegraphics[width=0.4\textwidth]{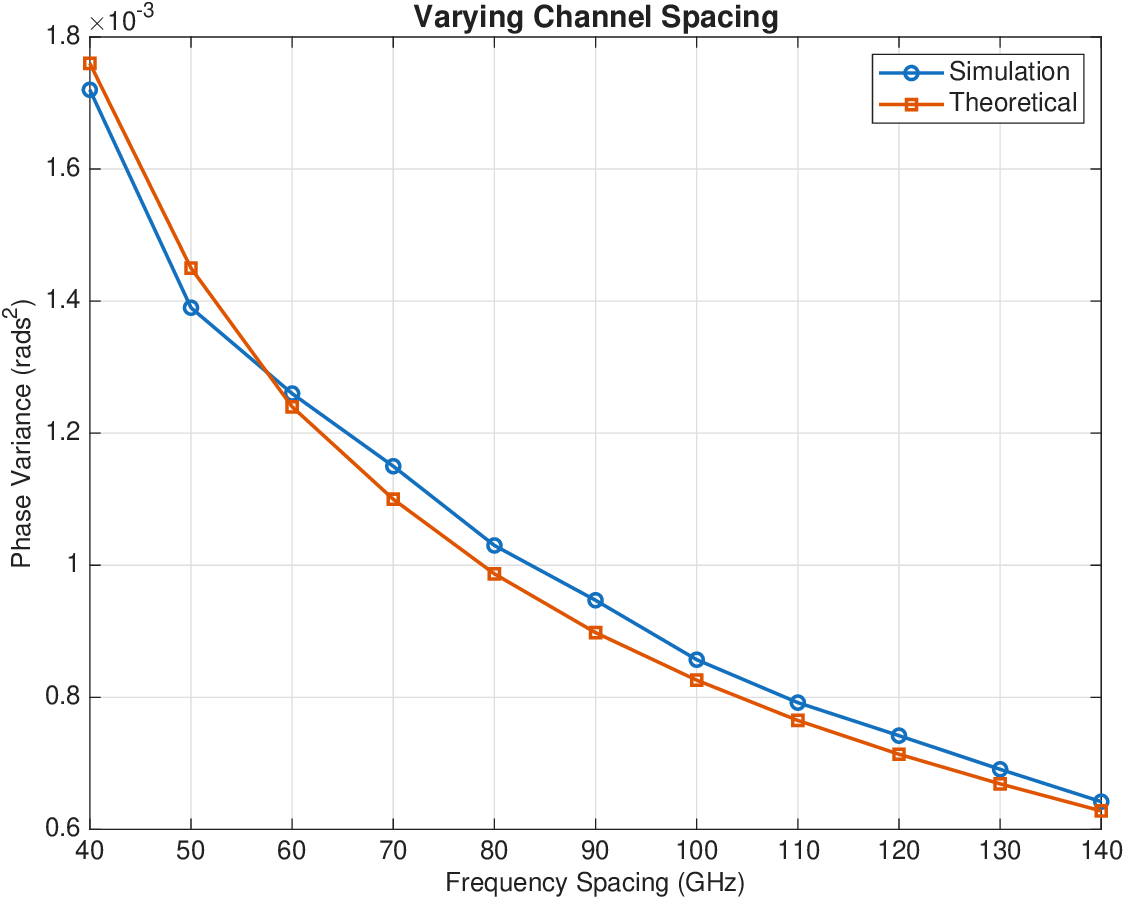}
        \label{fig:VaryingChannelSpacing}
    }
    \subfigure[]{
        \includegraphics[width=0.4\textwidth]{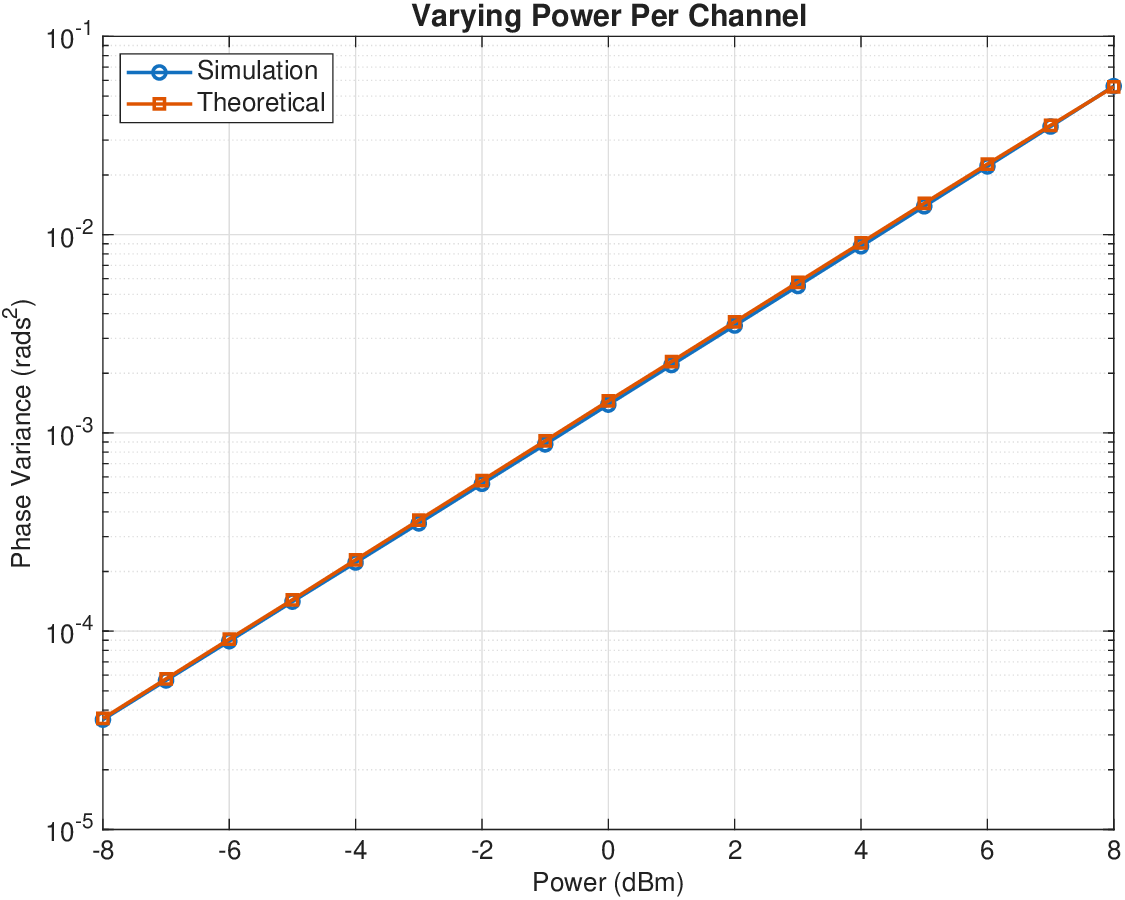}
        \label{fig:VaryingPower}
    }
    \caption{Comparison of average phase variance for a single subcarrier system against different parameters evaluated using Equation~(\ref{eq:AverageXPMPassBand}) with $K = 4/\pi$: (a) varying transmission distance (inserts: optical spectra at $800$ km and $2000$ km), (b) varying dispersion values (inserts: IF spectra at $2$ ps/nm/km and $16$ ps/nm/km), (c) varying channel spacing, and (d) varying channel power, and channel spacing of $50$ GHz. In all subfigures, the channel power (where not varied) is $1$ mW, the channel spacing (where not varied) is $50$ GHz, with a dispersion parameter (where not varied) of $16$ ps/nm/km, and distance (where not varied) of $800$ km.}
    \label{fig:VaryingPrameters}
\end{figure*}

\subsubsection{Application to Multi-Subcarrier Pumps}
Figure~\ref{fig:TenSpanXPM} further validates the model by applying it to a more complex two-subcarrier pump signal in a 10-span system. For this case, each subcarrier has a symbol rate of $R = 16$ GHz and a root-raised-cosine pulse shape with $\beta = 0.05$. The subcarrier spacing is set to $R \times (1+\beta) + 0.25$ GHz. The model correctly predicts the key spectral features: the broadened peaks (labeled "A") corresponding to the sinc-like link factor, and a spectral dip (labeled "B") caused by the frequency gap between the two subcarriers.

The sensitivity of the result to the integration range $\Delta\lambda$ was also investigated. The simulated phase variance was $1.264 \times 10^{-3}~\text{rad}^2$. The analytical calculation using Equation~(\ref{eq:AvgVar}) yielded $1.306 \times 10^{-3}~\text{rad}^2$, corresponding to a relative deviation of approximately 3.35\%. This result was achieved when integrating over the NB of each subcarrier. The integration was therefore performed over the Nyquist band of each subcarrier separately, rather than over the entire interval spanning both subcarriers and the gap between them. Including the excess bandwidth from the pulse-shaping roll-off ($\beta=0.05$) in the integration range (as illustrated in (A) of Figure~\ref{fig:1}) smears out the spectral dip "B" and provides no improvement in the total variance calculation. The DC component of the XPM term is excluded in all calculations, as it introduces only a static phase offset.

\subsection{XPM Model Validation}
It has been established that the model captures the XPM phase fluctuation's spectral characteristics. This section validates the model's ability to predict the total average phase variance, $\sigma_{\text{XPM}}^2$, in different system configurations. To do so, the phase variance calculated from Equation~(\ref{eq:AvgVar}) was compared against VPI simulation results. In Figure~\ref{fig:VaryingDistance}, the phase variance is shown to increase with distance. This is the expected result, as the vector summation of IFs (Equation~\ref{eq:ModifiedPhasor}) grows with the number of spans. The analytical model and simulation results are in good agreement for the first few spans. However, a discrepancy between the two emerges and increases beyond approximately 15 spans. This discrepancy is not a failure of the analytical model but rather a known limitation of the VPI simulation environment. In the simulation, demultiplexing the probe from the pump signal is not perfect. As shown in the inserts of Figure~\ref{fig:VaryingDistance} and clearly depicted in Figure~\ref{fig:OpticalSpectraPumpProbeOneSub}, the pump's sidebands grow with distance and begin to overlap with the probe's sidebands. This spectral overlap makes it impossible for the simulation to perfectly isolate the probe signal, leading to an inaccurate measurement of its phase variance. The analytical model, in contrast, assumes perfect probe isolation and is therefore not affected by this simulation artifact.

\begin{figure}[ht!]
    \centering
    \includegraphics[height=6cm, width=9cm]{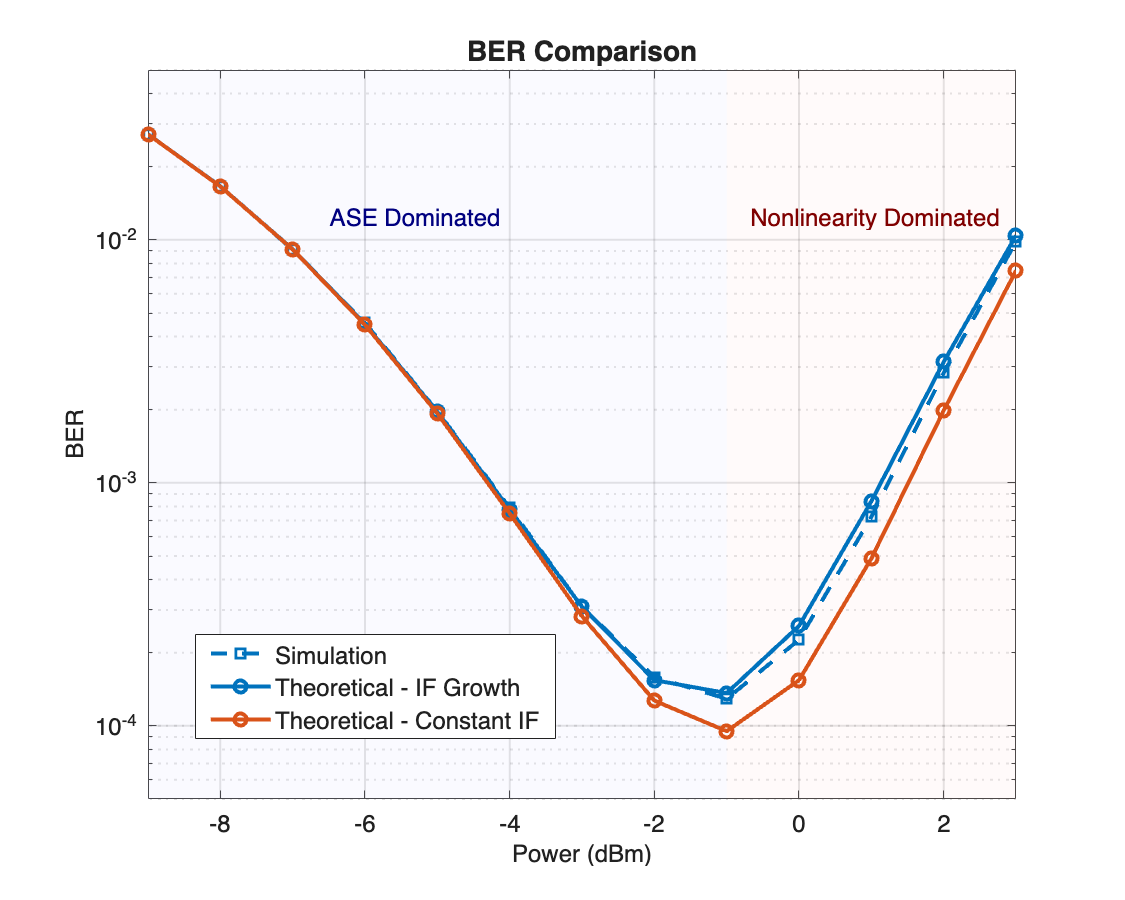}
    \caption{BER for a 16-QAM system after 30 spans, with two subcarriers spaced $50~\text{GHz}$ apart from another two subcarriers inducing XPM. The model incorporating IF growth closely matches simulation results, while assuming constant IF gives a lower BER under the same controlled radial-SNR comparison.}
    \label{fig:BER}
\end{figure}

In Figure \ref{fig:VaryingDispersion}, the calculated XPM phase variance decreases as the magnitude of the chromatic dispersion parameter increases. This well-known behaviour arises because low dispersion results in minimal walk-off between interacting channels, allowing phase shifts to accumulate coherently over a longer effective interaction length, which leads to a larger total phase variance. Conversely, higher dispersion induces significant walk-off, which rapidly decorrelates the phase contributions, limits the coherent accumulation, and results in a lower integrated phase variance.The analytical model presented in the figure was calculated using the incoherent-case approximation, $K = 4/\pi$. This model agrees well with the simulation results at high dispersion values, where the "incoherent sum" assumption holds. However, a notable discrepancy emerges at low dispersion values. This deviation occurs because the system is transitioning from the incoherent regime (where $K=4/\pi$ is appropriate) toward the coherent regime (where $K=1$ would be correct). In low-dispersion fibres, pulse broadening occurs more gradually, and as illustrated by the insets, the intensity spectral variation is much slower. This physical behaviour increases the coherence of the vector summation, diminishing the validity of the fixed $K=4/\pi$ approximation. Accordingly, the present model should be interpreted using the limiting statistical factors and the ratio-based framework discussed in Section~\ref{section:1}. For a system operating in the transition region, the relevant distribution of the phasor-sum ratio can be estimated for that system and substituted into $K=K_{\sigma}(QK_a)^2$. We therefore provide the bounds and the framework for computing the appropriate relative factor, while avoiding a universal interpolation formula because the intermediate distribution changes with dispersion, bandwidth, span count, and signal statistics.

The channel spacing (i.e., frequency spacing between pump and probe) also has a significant impact on the XPM phase variance, as depicted in Figure~\ref{fig:VaryingChannelSpacing}. The phase variance decreases as the channel spacing increases. This is because a larger frequency separation enhances the group velocity difference (walk-off) for a given fibre dispersion. This increased walk-off causes the signals in adjacent channels to pass through each other more rapidly, reducing their effective interaction time and consequently limiting the accumulation of XPM-induced phase noise.The analytical model captures this trend effectively, with a small discrepancy of 2-5\% compared to the simulation results. This minor deviation is attributed to the $K \approx 4/\pi$ factor (derived in Section~\ref{section:1}) being an \textit{average-case} approximation. While this average holds well across the integration, it is not the exact statistical correction for every specific frequency and parameter combination, leading to the small observed difference.

Finally, a quadratic relationship between the channel power and the phase variance is observed in Figure~\ref{fig:VaryingPower}, appearing as a $~$2:1 slope on the log-log plot (a 5~dB power increase causes a 10~dB variance increase). The analytical model and simulation results are in excellent agreement. This result strongly validates the Rayleigh amplitude assumption (Equation~\ref{eq:RayleighMeanPower}) used in the model's derivation, as it correctly predicts the XPM variance's dependence on pump power.

\subsection{BER Analysis}
The phase variance resulting from XPM distortions allows for direct Bit Error Ratio (BER) estimation. We utilize generalized closed-form expressions for M-QAM systems \cite{jafari2020bit,jafari2023generalized}, calculating the average BER by integrating the conditional error probability over a zero-mean Gaussian phase noise PDF with variance $\sigma_{\text{XPM}}^{2}$. The required Signal-to-Noise Ratio (SNR) is derived from the simulated Error Vector Magnitude (EVM). To separate amplitude perturbations from phase distortions, we isolate the radial SNR component ($\text{SNR}_{\text{Rad}}$) by decoupling the phase noise contribution from the total EVM-derived SNR \cite{shafik2006extended,ryu2003phase}. The EVM-derived SNR obtained directly from the received constellation contains the combined effect of radial amplitude noise and angular phase noise, and is therefore denoted as $\mathrm{SNR}_{\mathrm{Rad+PN}}$. Following the standard EVM--SNR relation \cite{shafik2006extended} and the small-phase-noise approximation used in phase-noise performance analysis \cite{ryu2003phase}, the inverse SNR contributions can be approximated as additive,
\begin{equation}
\mathrm{SNR}_{\mathrm{Rad+PN}}^{-1}\approx \mathrm{SNR}_{\mathrm{Rad}}^{-1}+\mathrm{SNR}_{\mathrm{PN}}^{-1}, \qquad \mathrm{SNR}_{\mathrm{PN}}^{-1}\approx \sigma_{\mathrm{XPM}}^2.
\end{equation}
Thus, the radial SNR used in the BER expression is obtained as
\begin{equation}
\mathrm{SNR}_{\mathrm{Rad}} \approx \frac{1}{\mathrm{SNR}_{\mathrm{Rad+PN}}^{-1}-\sigma_{\mathrm{XPM}}^2}.
\end{equation}

This extracted $\text{SNR}_{\text{Rad}}$, combined with the analytical variance $\sigma_{\text{XPM}}^{2}$, serves as the input for the BER model. For this evaluation, the CW probe is replaced by a 16-QAM signal (32 GHz bandwidth) identical to the pump. Both transmitters operate at 1 mW launch power with EDFA noise figures of 5 dB. Signals are demodulated via a coherent receiver without DSP-based distortion compensation.

Figure~\ref{fig:BER} illustrates the BER results, demonstrating strong agreement between the model and VPI simulations for the 16-QAM, two subcarrier system after 30 spans. Minor differences arise due to assumptions such as Gaussian distributed phase noise and the direct extraction of $E_b/N_0$ from simulations. The comparison between the evolving-IF and constant-IF assumptions is made at the same extracted radial SNR in order to isolate only the effect of the XPM phase variance. It should therefore not be interpreted as two independent launch-power simulations with identical nonlinear SNR. Under this controlled equal-$\mathrm{SNR}_{\mathrm{Rad}}$ comparison, neglecting IF growth predicts a smaller phase variance and consequently a lower BER in the nonlinear regime (above $-1$ dBm). This strongly highlights the importance of accurately accounting for phase variance in modeling.

\begin{figure}[hb!]
    \centering
    \includegraphics[height=6cm, width=9cm]{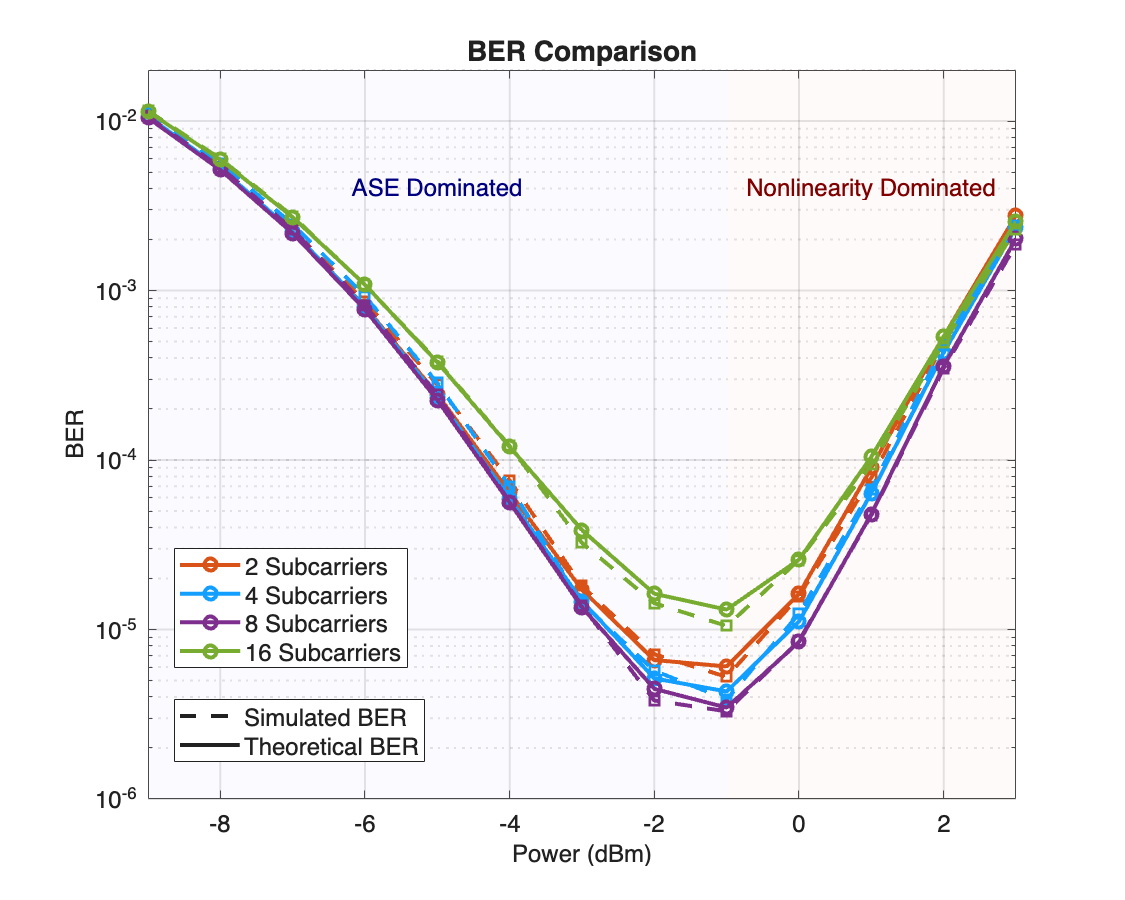}
    \caption{BER for a 16-QAM system at 20 spans for 3 different granular systems.}
    \label{fig:GranularBER}
\end{figure}

\begin{figure}[t]
    \centering
    \includegraphics[height=6cm, width=8cm]{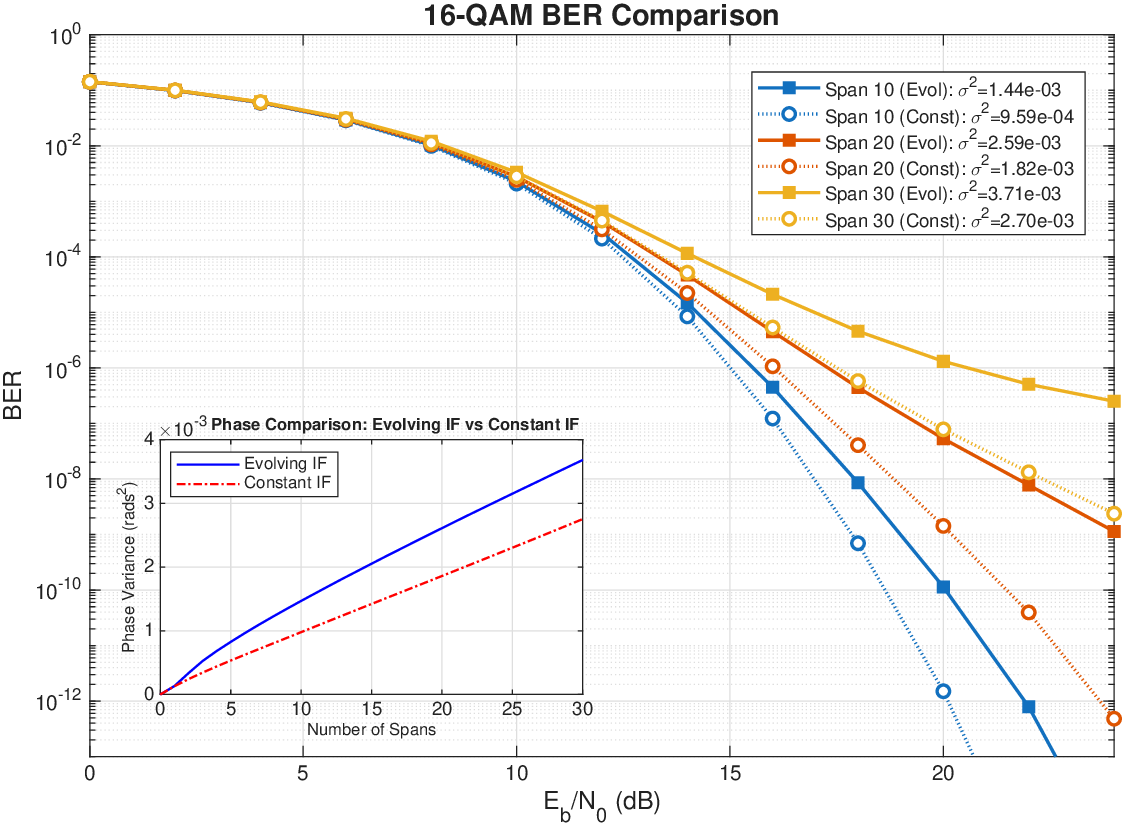}
    \caption{Predicted 16-QAM BER versus $E_b/N_0$ derived from the XPM-induced phase-variance model for a single-subcarrier inducing XPM on a single-subcarrier spaced by 50 GHz. Results are compared for evolving IF versus constant IF assumptions at 10, 20, and 30 spans. Inset: Accumulation of the modelled phase variance with span count for both IF assumptions; these variances directly determine the corresponding BER curves.}
    \label{fig:BERDifferIF}
\end{figure}

Figure~\ref{fig:GranularBER} presents the BER for a 16-QAM system under different granularity levels after 20 spans. The results confirm that the optimal granularity minimizes the phase variance, resulting in the lowest BER. However, phase noise is not the only factor affecting the BER; other noise sources, such as amplitude noise, also play a significant role. The model presented here focuses on the phase noise component, which is particularly relevant in systems where XPM-induced distortions are significant.

Figure~\ref{fig:BERDifferIF} extends the launch power comparison from Figures~\ref{fig:BER} and \ref{fig:GranularBER}  by plotting the theoretical 16-QAM BER \cite{jafari2020bit} against SNR per bit,  $E_b/N_0$. The results compare the ``evolving IF'' model (incorporating span-by-span growth) against the ``constant IF'' assumption for a pump-probe pair with 50\,GHz separation. For this figure, the same $E_b/N_0$ axis is used for both cases as a controlled comparison so that the impact of changing only the accumulated XPM phase variance can be observed. In a full launch-power comparison, the constant-IF and evolving-IF cases could also lead to different nonlinear SNRs. The inset displays the accumulated phase variances driving these BER curves. Two key trends emerge. First, the evolving IF model predicts significantly higher accumulated phase variance as the span count increases (inset, Figure~\ref{fig:BERDifferIF}), rendering the constant-IF assumption progressively less accurate and overly optimistic for long links. Second, the resulting BER discrepancy is most pronounced in the high-SNR ``waterfall'' region, where performance is limited by residual phase noise rather than additive noise.

\section{Conclusion}\label{Conclusion}
Since XPM is dependent on the fluctuations in intensity resulting in the IF spectra to grow, the XPM equation was modified to incorporate this IF growth that has been modelled in \cite{Prasad2025intenstiy} while preserving the conventional understanding of XPM to approximate the XPM on a probe induced by a pass band pump signal. The results suggested indeed, the IF growth affects the XPM distortions. It also suggested that reducing intensity fluctuations especially around the lower frequencies could reduce the main lobe closer to the DC component that could significantly reduce the phase variance induced by the XPM as shown through the phase fluctuation spectra. The phase spectrum model was validated against the VPI simulation results and was able to emulate the phase fluctuation spectra of the XPM component. The model was also able to capture the phase variance of the XPM component on a probe with a small discrepancy between the simulation and analytical model under different system parameters. Using these phase variance results, it was also demonstrated the full system performance of average BER of different granular systems could be determined through the average phase variance of the XPM component using a BER equations derived for a system under the influence of phase noise.

Although the analytical structure of the model is general, the numerical validation presented in this work is limited to SSMF-type dispersion values, 80-km spans, and 16-QAM signals. The main objective of the present paper is to demonstrate that dispersion-induced IF growth must be included in the linear computation of XPM spectra, phase variance, and BER; it is not intended to exhaustively validate every fibre type, modulation format, or shaped-signal distribution. Other fibre types, such as non-zero dispersion-shifted fibres, different span lengths, different symbol rates, and higher-order QAM formats can change the walk-off, the IF spectral evolution, and the coherent-to-incoherent transition represented by $K$. The model should therefore be revalidated, or the statistical factor recalibrated, when it is applied far outside the parameter range considered here.

Future extensions of this work may include a dual-polarization Manakov-form implementation, validation for non-zero dispersion-shifted fibres and alternative span lengths, and application to higher-order modulation formats, shaped constellations, and non-contiguous multi-subcarrier spectra. A further useful extension would be to use the ratio-based framework developed here to evaluate a parameter-dependent statistical factor $K$ for transition-region systems, where the relevant phasor-sum distribution changes with dispersion, bandwidth, span count, and modulation statistics.

\section*{Acknowledgment}
The authors wish to express sincere gratitude to Professor Arthur Lowery for his invaluable guidance, insightful discussions, and constructive feedback throughout the course of this research. This work was supported by the Faculty of Engineering Publication Award, Monash University.

\appendix
\section{Derivation of Amplitude Ratio Bounds}
\label{app:bounds}

This appendix derives bounds for the expectation ratio $Q$, referenced in the analysis of the incoherent pump model. We consider the random phasor sum defined as:
\begin{equation}
    \upsilon = \sum_{k=1}^N a_k e^{-j\phi_k},
\end{equation}
where the amplitudes $a_k$ are i.i.d. random variables with mean $\mu$ and variance $\sigma^2$, and the phases are $\phi_k = C(k-1)$. We define the deterministic geometric factor as $R = \left|\sum_{k=1}^N e^{-j\phi_k}\right|$. By linearity of expectation, the denominator of the ratio is:
\begin{equation}
    |\mathbb{E}[\upsilon]| = \left| \mu \sum_{k=1}^N e^{-j\phi_k} \right| = \mu R.
\end{equation}

To bound $Q$, we first apply Jensen’s inequality ($\mathbb{E}[|X|] \ge |\mathbb{E}[X]|$) to establish the lower bound $Q \ge 1$. For the upper bound, we utilize the RMS inequality $\mathbb{E}[|\upsilon|] \le \sqrt{\mathbb{E}[|\upsilon|^2]}$. For independent amplitudes, the second moment is the sum of individual variances plus the squared magnitude of the mean vector:
\begin{equation}
    \mathbb{E}[|\upsilon|^2] = \sum_{k=1}^N \mathrm{Var}(a_k) + |\mathbb{E}[\upsilon]|^2 = N\sigma^2 + \mu^2 R^2.
\end{equation}

Substituting these results into the ratio definition $Q = \mathbb{E}[|\upsilon|] / |\mathbb{E}[\upsilon]|$ yields the general bounds:
\begin{equation}
    1 \le Q \le \frac{\sqrt{N\sigma^2 + \mu^2 R^2}}{\mu R} = \sqrt{1 + \frac{N \sigma^2}{\mu^2 R^2}}.
\end{equation}

We consider the specific case of Rayleigh-distributed amplitudes, where the variance-to-mean-squared ratio is $\sigma^2/\mu^2 = (4-\pi)/\pi$. Additionally, averaging the geometric factor over a full phase period $C \in [0, 2\pi]$ results in $\langle R^2 \rangle = N$. Substituting these specific values provides the average-case upper limit:
\begin{equation}
    Q_{\text{avg}} \le \sqrt{1 + \frac{N(4-\pi)}{\pi N}} = \sqrt{\frac{4}{\pi}} \approx 1.128.
\end{equation}
This result confirms that the expectation ratio is tightly constrained to the interval $1 \le Q \le \sqrt{4/\pi}$.

\bibliographystyle{elsarticle-num}
\bibliography{Ref}

\end{document}